\documentclass[%
	twocolumn,
	superscriptaddress,
	amsmath,amssymb,
	aps,
	prl,
	letterpaper,
	floatfix
	]{revtex4-1}
\usepackage{color,soul}

\allowdisplaybreaks
\usepackage{graphicx}
\usepackage{dcolumn}
\usepackage{bm}
\usepackage{mathdots}
\usepackage[dvipsnames]{xcolor}

\soulregister\cite7
\soulregister\ref7

\begin{document}


\title{Investigation of two-frequency Paul traps for antihydrogen production}

\author{Nathan Leefer}
\email{naleefer@berkeley.edu}
\affiliation{Helmholtz-Institut Mainz, Mainz 55128, Germany}
\affiliation{Department of Physics, University of California at Berkeley, Berkeley, CA 94720}
\author{Kai Krimmel}
\affiliation{Helmholtz-Institut Mainz, Mainz 55128, Germany}
\affiliation{QUANTUM, Institut f\"ur Physik, Johannes Gutenberg-Universit\"at Mainz, Mainz 55128, Germany}
\author{William Bertsche}
\affiliation{University of Manchester, Manchester M13 9PL, UK }
\affiliation{The Cockcroft Institute, Daresbury Laboratory, Warrington WA4 4AD, UK}
\author{Dmitry Budker}
\affiliation{Helmholtz-Institut Mainz, Mainz 55128, Germany}
\affiliation{QUANTUM, Institut f\"ur Physik, Johannes Gutenberg-Universit\"at Mainz, Mainz 55128, Germany}
\affiliation{Department of Physics, University of California at Berkeley, Berkeley, CA 94720}
\affiliation{Nuclear Science Division, Lawrence Berkeley National Laboratory, Berkeley, CA 94720}
\author{Joel Fajans}
\affiliation{Department of Physics, University of California at Berkeley, Berkeley, CA 94720}
\author{Ron Folman}
\affiliation{Department of Physics, Ben-Gurion University of the Negev, Be'er Sheva 84105, Israel}
\author{Hartmut H\"affner}
\affiliation{Department of Physics, University of California at Berkeley, Berkeley, CA 94720}
\author{Ferdinand Schmidt-Kaler}
\affiliation{Helmholtz-Institut Mainz, Mainz 55128, Germany}
\affiliation{QUANTUM, Institut f\"ur Physik, Johannes Gutenberg-Universit\"at Mainz, Mainz 55128, Germany}

\date{\today}

\begin{abstract}
Radio-frequency (rf) Paul traps operated with multifrequency rf trapping potentials provide the ability to independently confine charged particle species with widely different charge-to-mass ratios. In particular, these traps may find use in the field of antihydrogen recombination, allowing antiproton and positron clouds to be trapped and confined in the same volume without the use of large superconducting magnets. We explore the stability regions of two-frequency Paul traps and perform numerical 
simulations of small samples of multispecies charged-particle mixtures of up to twelve particles that indicate the promise of these traps for antihydrogen recombination.
\end{abstract}

\pacs{Valid PACS appear here}

\maketitle

\section{\label{sec1}Introduction}
The measurable properties of hydrogen ($H$) and antihydrogen ($\bar{H}$) atoms are expected to be identical as postulated by the combined charge (C), parity (P), and time (T) reversal symmetry~\cite{Luders1957}. One of the most promising tests of this symmetry is the precise comparison of the optical and microwave spectra of hydrogen and antihydrogen. The spectrum of hydrogen has been extensively studied~\cite{Parthey2011, Hardy1979}, but precise measurements for antihydrogen are complicated by the small quantities of $\bar{H}$ available and the technical complexity of the experimental apparatus~\cite{Amole2012}. The efficient production and trapping of cold and neutral antimatter systems is therefore a topic of great interest.

Production of antihydrogen requires the ability to trap antiprotons and antielectrons (positrons) in the same volume. The state of the art is dominated by Penning traps, where a constant homogeneous magnetic field and inhomogeneous static electric field allow for confining particles of mass $m$ and charge $Q$.  The ALPHA experiment~\cite{Andresen2010,Andresen2011} and the ATRAP experiment ~\cite{Storry2004,Gabrielse2012} rely on a variation of a Penning trap for initial particle confinement. Penning traps have the advantage of robust trapping for a wide range of charge-to-mass ratios, while also facilitating a high charge density of positrons for efficient three-body recombination. A large trap volume and superconducting magnet creates a high magnetic trap depth (~1 K) for the resulting neutral antiatoms. A limitation, however, is the inability to trap the oppositely charged particles in equilibrium in the same volume due to the use of a DC potential for confinement along the axial trap direction. Recombination is achieved by injecting antiprotons into the positron cloud~\cite{Amole2013}. The resulting antiatoms are typically created with energy above the magnetic trap depth, and most antiatoms are lost during recombination. Typical yields in the ALPHA apparatus are several trapped antiatoms per attempt every $\approx$15 minutes~\cite{Amole2014,Amole2014a}. The ASACUSA experiment has pursued an alternative to spectroscopy on trapped atoms with a CUSP trap~\cite{Mohri2003}, which uses an anti-Helmholtz field to generate a beam of spin polarized antihydrogen for eventual microwave spectroscopy atoms~\cite{Enomoto2010,Kuroda2014}.

A solution to the problem of equilibrium charge overlap was previously explored in a hybrid Penning-Paul trap~\cite{Walz1995}. In that work a magnetic field and DC potential of a Penning trap confined protons and the radio-frequency Paul trap potential compensated the axial DC potential for electrons. The method still relied on a strong magnetic field for radial confinement, and to our knowledge this technique has not been continued or extended to antimatter systems.

Two-dimensional confinement of electrons has been achieved in planar devices~\cite{Hoffrogge2011}, but there are no reports on confinement of ions with high charge-to-mass ratio $Q_m:=Q/m$ in a three-dimensional trap. In order to achieve such three-dimensional confinement, the stability parameters of electrons --- or in our case positrons --- would need to be worked out. For antihydrogen formation, we approach the problem of simultaneous three-dimensional particle confinement of antiprotons and positrons with the idea of a two-frequency Paul trap. This trap design is aimed to combine the stability parameters of both particles and would also allow for charge overlap inside the trap.

A Paul trap provides a dynamical trapping potential in all three space directions, and works for positive and negative charges equally well. The problem arises from the vastly different charge-to-mass ratio of antiprotons ($\bar{p}$) and positrons ($\bar{e}$). The stability of a Paul trap is characterized by dimensionless stability parameters $a$ and $q$~\cite{Seriesa}, which are related to the static and dynamic amplitudes, respectively, of the confining potential. Both parameters scale linearly with the charge-to-mass ratio, $Q_m$. A is stable for $0<q<0.9$ in case of $a\approx 0$, with optimal trapping achieved around $q = 0.5$. A trap optimized for trapping antiprotons will have an effective $q \approx 900$ for positrons and is fully \mbox{unstable.}

A Paul trap optimized for positrons with large $Q_m$ is theoretically stable for antiprotons, but suffers from poor equilibrium charge overlap. Particle confinement is characterized by the pseudopotential $U \propto m (a+q^2) \Omega^2 r^2$, where $m$ is the particle mass, $\Omega$ is the frequency of the trap potential, and $r$ is the distance from the trap center~\cite{Goldman2014}. If antiprotons and positrons confined in the same region thermalize due to the Coulomb interaction and $a\approx 0$, the characteristic cloud radius of antiprotons will be a factor of $\sqrt{m_{\bar{p}}/m_{\bar{e}}} \approx 45$ larger due to the dependence of $q$ on $Q_m$. The larger cloud radius of antiprotons will also make them more susceptible to anharmonicities of the trapping potential. We note that the ASACUSA collaboration reported work for several years on a large-volume, superconducting resonant-cavity Paul trap for antihydrogen production~\cite{ASACUSA2007,*ASACUSA2008,*ASACUSA2009,*ASACUSA2010,*ASACUSA2011}. More recent reports indicate the intention to use this trap for spectroscopy of antiprotonic helium, $\bar{p}$He$^+$~\cite{ASACUSA2014}.

\begin{figure}[t]
\includegraphics[width=\linewidth]{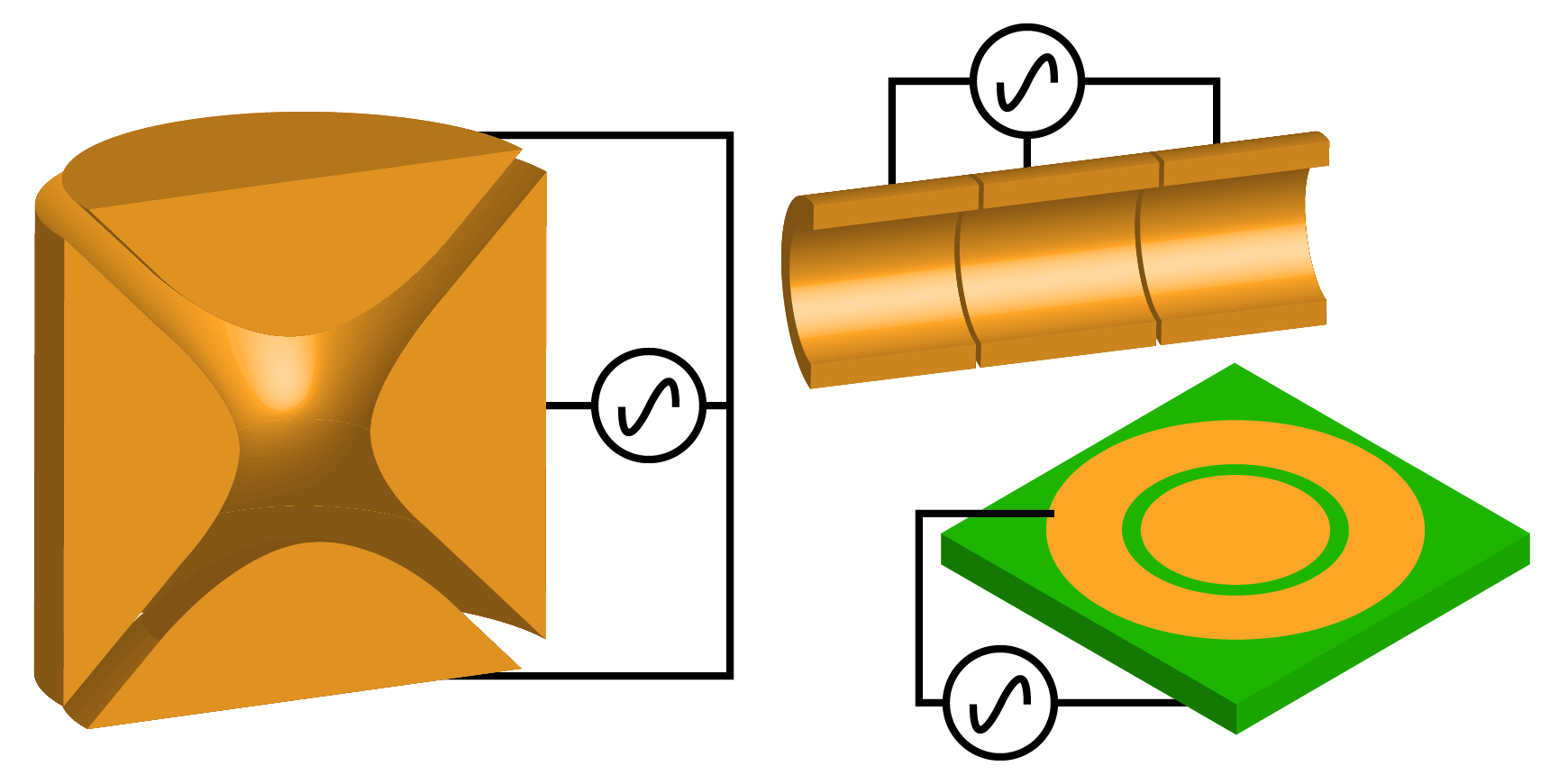}
\caption{\label{fig1} Survey of various trap geometries that can realize the potential indicated in Eq.~\eqref{eq1}. Particularly interesting are planar all-rf Paul traps indicated by the geometry in lower right. Such a geometry is suitable for miniaturization with modern atom chip technology~\cite{Reichel2011, Keil2016}. Atom chip technology may then also support deep traps for the produced neutral antihydrogen.}
\end{figure}

In this paper we discuss features of a two-frequency Paul trap with an 
infinite, perfect quadrupole potential
that allows simultaneous confinement of antiprotons and positrons \textit{and} allows the antiproton and positron cloud sizes to be matched. Trap frequencies are chosen such that positrons are confined by the high-frequency component of the trap potential and protons are primarily confined by the low-frequency component. This allows the pseudopotentials for antiprotons and positrons to be adjusted independently. Our work was partially inspired by the preliminary discussion of two-frequency Paul traps in Ref.~\cite{Trypogeorgos2013}.

\section{\label{sec2}Two-frequency Paul trap}
The quadrupole potential of a two-frequency Paul trap takes the form 
\begin{equation}\label{eq1}
V(t,\mathbf{r}) = (V_0 + V_1 \cos{\Omega_1 t} + V_2 \cos{\Omega_2 t})\frac{(x^2 + y^2-2 z^2)}{2 r_0^2},
\end{equation}
where $r_0$ is a geometric scale for the trap. We will choose frequencies such that the fraction $\Omega_2/\Omega_1$ is a number $\eta \geq 1$. The potential can be created by a system of hyperbolic electrodes with cylindrical symmetry, or approximated by more practical geometries as indicated in Fig.~\ref{fig1}.

In the initial discussion only motion along the x-direction is considered. From the symmetry of the potential these results will also hold for the y-direction, and may be extended to the z-direction by scaling stability parameters by a factor of $-2$. The equation of motion for a charged particle in the potential of Eq.~\eqref{eq1} can be written
\begin{equation}\label{eq2}
\ddot{x}(\tau) + (a - 2 q_1 \cos{2 \eta^{-1} \tau} - 2 q_2 \cos{2 \tau})x(\tau) = 0,
\end{equation}
where $\tau = \Omega_2 t/2$ is a normalized time, 
\begin{align}
q_{1,2} &= -2 Q_m \frac{V_{1,2}}{\Omega_2^2 r_0^2},\\
a &= 4 Q_m \frac{V_0}{\Omega_2^2 r_0^2}
\end{align}
are low-frequency ($q_1$), high-frequency ($q_2$), and DC ($a$) Mathieu parameters. The time derivative indicated by $\ddot{x}(\tau)$ is with respect to the time variable $\tau$. Equation~\eqref{eq2} is a specific example of a Hill differential equation: a second order, linear differential equation with periodic coefficients.

\subsection{\label{subsec1}Qualitative discussion}
Setting $q_1 = 0$ recovers the well known Mathieu equation for a single-frequency trap. A trapped particle undergoes high-frequency motion at multiples of the trap drive frequency, $\Omega_2$, in addition to a slow macromotion at a secular frequency of $\omega^2 = (1/4)(a + q_2^2/2)\Omega_2^2$. In an optimal trap $a\approx 0$, and the secular frequency is $\omega \approx q_2 \Omega_2/(2\sqrt{2})$~\cite{Seriesa}. If we operate the trap at $q_1 \approx 0,\,q_2 \approx 0.5$ we can choose the ratio of trap drive frequencies, $\eta$, large enough so that the secular oscillation frequency $\omega \gg \Omega_1$. In this regime it is possible to treat a non-zero $q_1$ as a slowly varying DC term in addition to $a$.

We now consider Eq.~\eqref{eq2} from the perspective of two charged particles, $A$ and $B$, with opposite charges and masses $m_A < m_B$. To facilitate the discussion we introduce the notation $q_{1,2}^{A,B}$ and $a^{A,B}$ to distinguish trap parameters for the light particle, $A$, and the heavy particle, $B$. An important observation is that $q_{1,2}^B = (m_A/m_B)q_{1,2}^A$. The same relationship holds for $a$, although this DC parameter will be set to zero for most of the manuscript.

\begin{figure}[t]
\includegraphics[width=\linewidth]{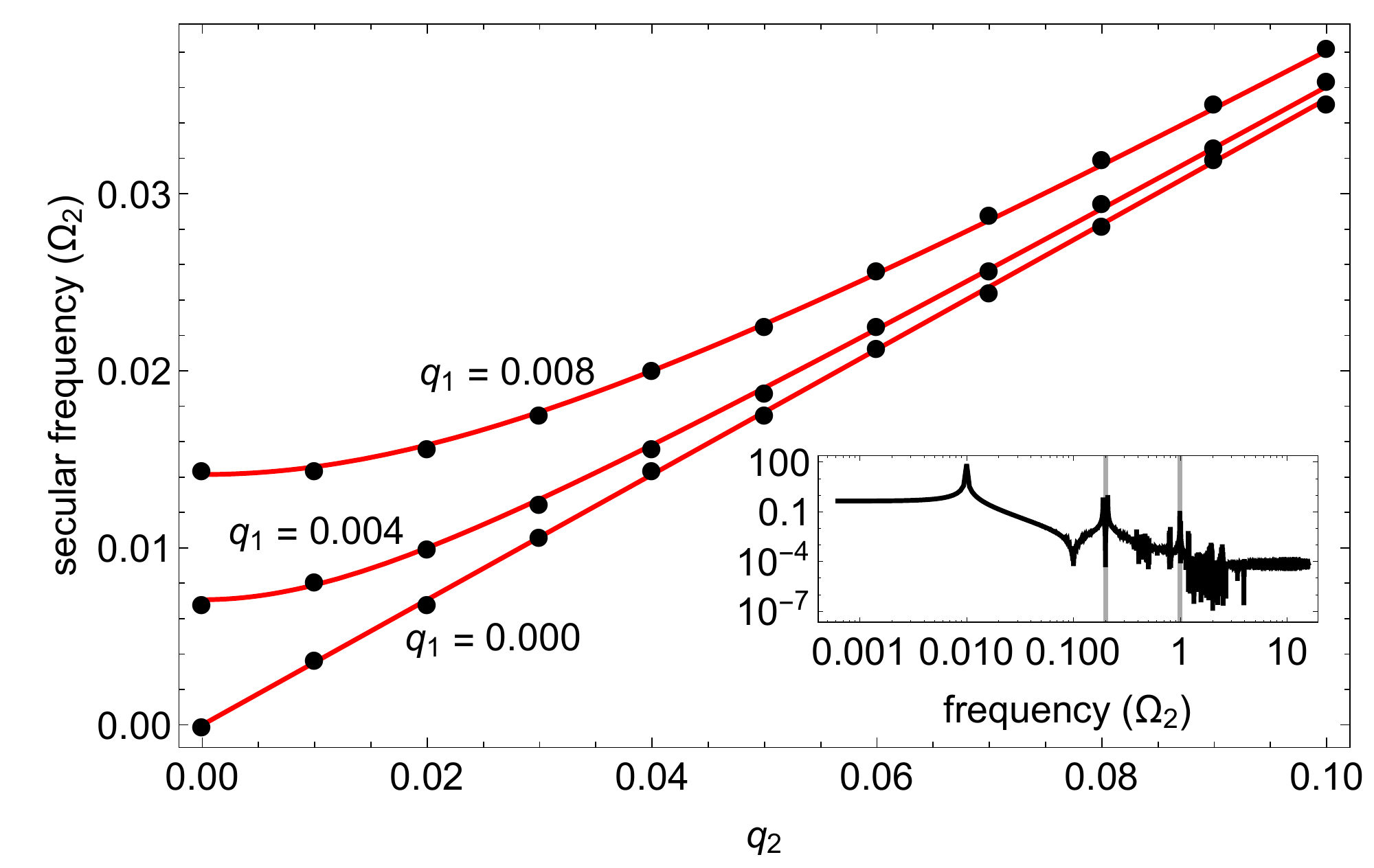}
\caption{\label{fig2} Scaling of the secular frequency as a function of $q_2$ for $B$ in units of $\Omega_2$, for $\eta = 5$. The values of $q_1$ for each curve are indicated on the plot. The value of $a$ was zero for all calculations. The solid lines indicate the expected secular frequency extracted from the pseudopotential of Eq.~\eqref{eq4}. Filled circles are estimated from a fast Fourier transform (FFT) of the numerical integration of Eq.~\eqref{eq2}. The inset shows the calculated FFT used to extract secular frequencies for $q_1 = 0.004,q_2 = 0.02$. Vertical lines indicate the driving frequencies. The lowest-frequency peak is the secular frequency. 
We do not include Coulomb interaction here but do so in Eq.~\ref{eq18}.
}
\end{figure}

If the high-frequency confinement is optimized for the lighter particle $A$ and for $\eta < m_B/m_A$, the secular oscillation frequency of the heavy particle $B$ due to $q_2^B$ will be slower than $\Omega_1$. Therefore we must also consider the dynamic pseudo potential of $q_1^B$ for $B$. We may write the effective potentials experienced by each particle as~\cite{Seriesa}

\begin{figure}[t]
\includegraphics[width=\linewidth]{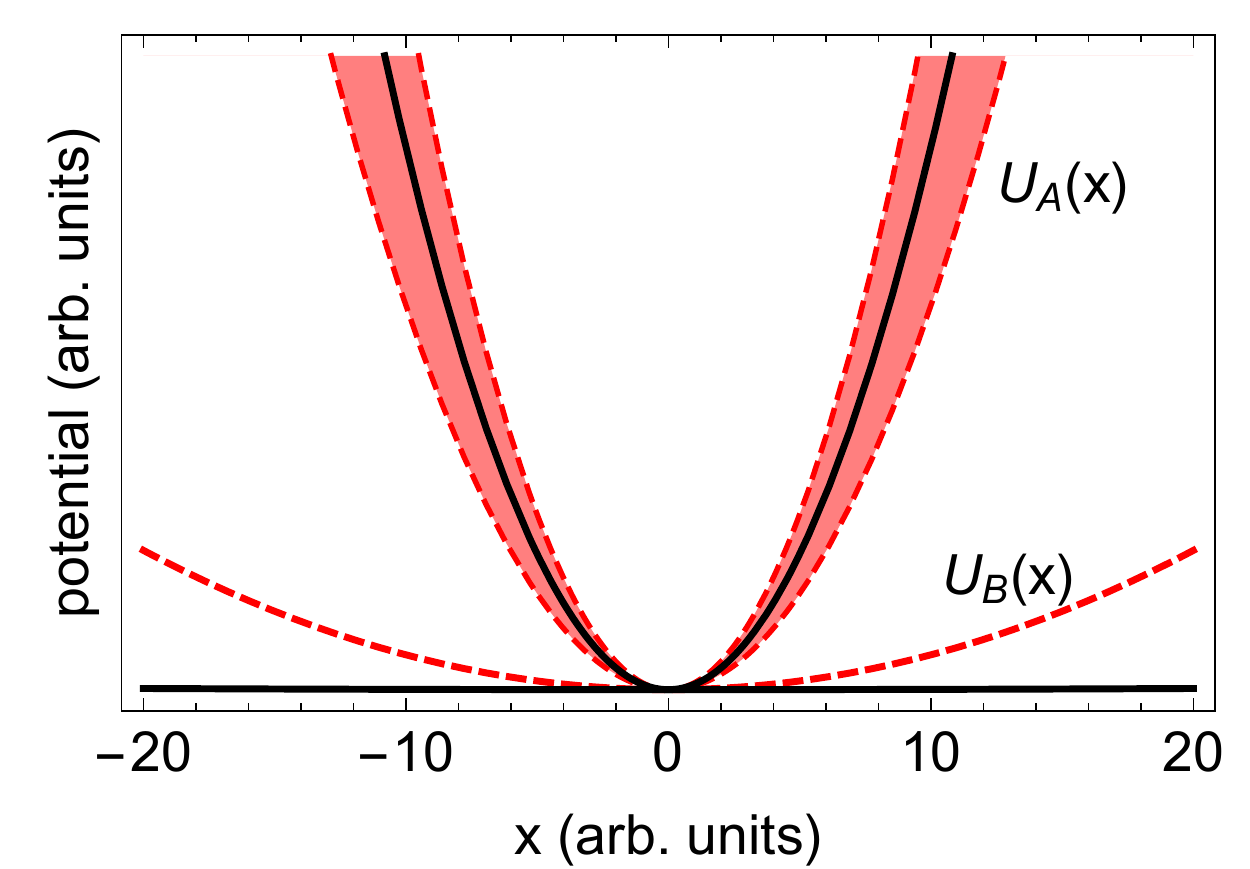}
\caption{\label{fig3} Sketch of pseudopotentials for particles $A$ and $B$ assuming a frequency ratio of $\eta = 200$ and mass ratio, $m_B/m_A \approx 1836$ (matching that of positrons and antiprotons). Parameters were $q_2^A = 0.37$, $q_1^A = 0$ (solid lines) and $q_2^A = 0.37$, $q_1^A =  0.02$ (dashed lines). The small perturbation of $U_A(x)$ by $q_1^A$ oscillates with frequency $\Omega_1$ as indicated by the shaded band and will average to zero.}
\end{figure}

\begin{samepage}
\begin{align}\label{eq4}
U_A(x) &= \frac{1}{8}m_A\left(a^A + \frac{(q_2^A)^2}{2}\right)\Omega_2^2x^2,\\
U_B(x) &= \frac{1}{8}m_B\left(a^B + \frac{(q_1^B)^2}{2}\eta^2 + \frac{(q_2^B)^2}{2}\right)\Omega_2^2x^2,
\end{align}
\end{samepage}
where we typically choose $\eta$ to be large enough that $q_1^A$ is close enough to zero that its effect on particle $A$ can be ignored. The validity of the pseudopotential approximation is discussed in detail in Refs.~\cite{Kaufmann2012, Landa2012a}. The pseudopotential for particle $B$ in Eq.~\eqref{eq4} can be arrived at by alternately considering the limiting cases where $q_{1,2}^{B}$ go to zero. When $q_2^{B}=0$, Eq.~\eqref{eq2} can be rewritten with another time transformation $\tau \rightarrow \tau'= \Omega_1 t/2 $ to obtain the $\eta^2$ factor multiplying $q_1^B$. We confirmed the scaling of these pseudopotentials by numerical determination of the secular frequency, shown in Fig.~\ref{fig2}.  Equation~\eqref{eq4} also illustrates the ability of a two-frequency trap to create overlapping yet independent potential wells for two charged particles with an appropriate choice of trap parameters and frequency ratio, visually demonstrated in Fig.~\ref{fig3}. This opens the possibility for combining and separating different species of charged particles with high efficiency.

In the remainder of the manuscript we discard the superscript notation. Where relevant, it is assumed that $q_{1,2}$ and $a$ refer to trap parameters for the \textit{lighter} charged particle.


\begin{figure*}[t]
 \includegraphics[width=\textwidth]{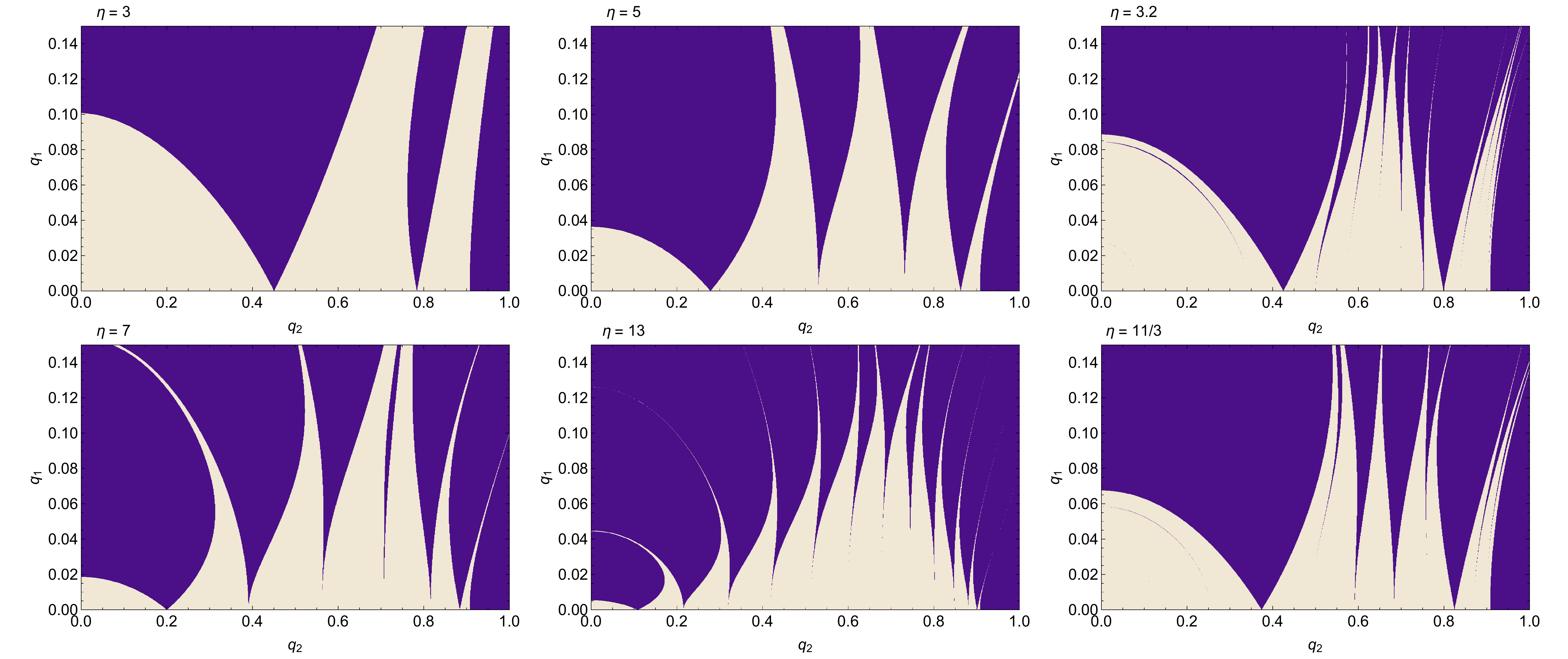}
 \caption{\label{fig5} Stability diagram of a two-frequency Paul trap for $q_1$ and $q_2$. Lighter shading indicates stable operating parameter regions. Frequency ratios, $\eta$, are indicated in the upper-left corner of each plot.}
\end{figure*}

\subsection{Floquet Theory}
To determine the stability of a two-frequency trap we define the vector $\mathbf{u}(\tau) = [x(\tau),\dot{x}(\tau)]$. Equation~\eqref{eq2} may then be written in matrix form
\begin{equation}\label{eq6}
\dot{\mathbf{u}}(\tau) = \mathbf{P}(\tau) \cdot \mathbf{u}(\tau),
\end{equation}
where
\begin{equation}\label{eq7}
\mathbf{P}(\tau) = \begin{pmatrix} 0 & 1\\ 
				(a - 2 q_1 \cos{2\eta^{-1}\tau}-2q_2\cos{2\tau}) & 0
        		\end{pmatrix}.
\end{equation}
If $\eta$ is a rational number it can be represented as an irreducible fraction $m/n$, where $m$ and $n$ are both integers and $m \ge n$. In this case the matrix $\mathbf{P}(\tau)$ has periodicity $T = m \pi$ such that $\mathbf{P}(\tau + T) = \mathbf{P}(\tau)$.

General closed-form solutions of Eq.~\eqref{eq6} do not exist, however it is possible to use Floquet theory to make statements about the existence of bound solutions for particular values of equation parameters. The existence of bound solutions implies a stable trap.

A discussion of Floquet theory may be found in most differential-equations texts, such as Ref.~\cite{Jordan2007}, and application of Floquet theory to Paul traps in Refs.\cite{Kaufmann2012, Landa2012a, Landa2012b}. Here we simply state that the boundary of stability regions may be found by identifying parameters for which the solution $x(\tau)$ has periodicity $T$ or $2T$. A general solution with period $2T$ contains all solutions with period $T$, so we choose

\begin{equation}\label{eq8}
x(\tau) = \sum_k c_k e^{i \frac{k}{m}\tau},
\end{equation}
where the sum over $k$ extends from $-\infty$ to $+\infty$. Equation~\eqref{eq8} and Eq.~\eqref{eq7} lead to the identity

\begin{samepage}
\begin{align}\label{eq9}
\sum_k\Bigg[\left( a - \frac{k^2}{m^2}\right)& c_k - q_1 (c_{k-2 n}+c_{k+2 n})-\notag\\
&q_2(c_{k - 2 m} + c_{k+2 m})\Bigg] e^{i \frac{k}{m}\tau} =0.
\end{align}
\end{samepage}
The only way for Eq.~\eqref{eq9} to hold for all $\tau$ is if each element of the sum satisfies this relation independently. Equation~\eqref{eq9} may therefore be summarized as a matrix equation

\begin{equation}\label{eq10}
\mathbf{D} \cdot \begin{pmatrix}
\vdots\\
c_{k-1}\\
c_k\\
c_{k+1}\\
\vdots
\end{pmatrix} = 0,
\end{equation}
where the elements of the infinite matrix $\mathbf{D}$ are given by 
\begin{multline}\label{eq11}
D_{ij} = \left(\eta^2a - \frac{k^2}{m^2}\right)\delta_{i,j} - q_1( \delta_{j,j-2n} + \delta_{j,j+2n}) -\\  q_2( \delta_{j,j-2m} + \delta_{j,j+2m}).
\end{multline}

\subsection{Stability diagrams}
Equation~\eqref{eq10} is equivalent to the statement 
\begin{equation}\label{eq12}
\det(\mathbf{D}) = 0.
\end{equation} 
Stable trap operating parameters can be identified by finding parameters that satisfy Eq.~\eqref{eq12}. Although $\mathbf{D}$ is an infinite matrix, a matrix of size $(10m+1) \times (10m + 1)$ centered around $D_{00}$ was found to be a sufficient approximation for evaluating stability. We find that parameters where $\det{(\mathbf{D})}>0$ correspond to stable traps, and parameters where $\det{(\mathbf{D})}<0$ correspond to unstable traps. This means only $\det{(\mathbf{D})}$ needs to be evaluated, and Eq.~\eqref{eq12} does not need to be solved exactly. Using larger matrices changes the stable area by less than 0.1\%. The matrix evaluated can still be large, but most elements are zero and programs such as \textit{Mathematica} or MATLAB have efficient tools for computations with sparse matrices. 

The stability diagrams in $q_1, q_2$ space are shown in Fig.~\ref{fig5} for integer and rational frequency ratios. These diagrams show a structure of unstable resonances that increase in density with the frequency ratio. Near the $q_2$ axis these unstable features correspond to a parametric resonance condition between the secular oscillation frequency $\omega$ of the particle due to $q_2$ and the frequency $\Omega_1$. These resonances become infinitely thin, but extend all the way to the $q_2$ axis. For large frequency ratios this structure indicates that a damping mechanism will be necessary for long-term stable operation of a two-frequency trap. The stability region for rational numbers shares general features with the closest integer ratio diagrams, with a significantly denser resonance structure as can be seen in Fig.~\ref{fig5}.

It is important to note that stability for the light particle does not guarantee stability for the heavy particle. Stability calculations can easily be evaluated for both particles, however simultaneous stability can be obtained with a general guideline. Revisiting particles $A$ and $B$: if $m_B/m_A \gg \eta$, stable values of $q_2$ and $q_1$ for the light particle are also stable for the heavy particle if $q_1 \eta^2 m_A/m_B < 0.9$. This can be seen by setting $q_2 = 0$ in Eq.~\eqref{eq2} for the heavy particle, and recovering the regular Mathieu equation with the transformation $\tau = \tau' \eta$. 

\section{\label{sec3}Numerical simulations}
In the previous section we require the determinant of the matrix $\mathbf{D}$ in Eq.~\eqref{eq10} to be zero. With this method we are forced to approximate an infinite matrix with a large but finite representation.

\begin{figure}[t]
\includegraphics[width=\linewidth]{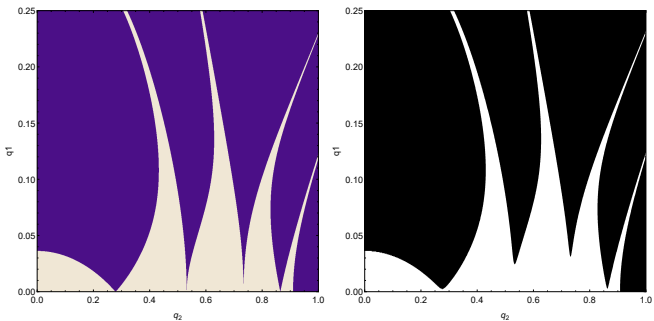}
\caption{\label{fig6}
A direct comparison of the a) matrix determinant and b) numerically calculated stability diagrams for $\eta=5$. The unstable resonances are finely resolved in the left diagram and extend all the way to the $q_2$-axis, while in the diagram to the right the stability arms merge before reaching the $q_2$-axis. For further comparison between the two methods, the analytic calculation took $~15$~s and the numerical calculation took $9700$~s, a difference of nearly three orders of magnitude. Regarding the scaling of the computation time, both methods take significantly longer for larger $\eta$.} 
\end{figure}

Another issue of the matrix determinant solution is that we are bound to rational frequency ratios $\eta$, which means that in practice an irrational frequency ratio can only be approximated by rounding to a nearby rational number. While the matrix method works for any rational number in principle, even short decimal numbers require the evaluation of impractically large matrices. For example, finding stable regions for $\eta = 3.14159$ would require the evaluation of a $314160\times314160$ matrix for comparable accuracy to the stability diagrams in Fig.~\ref{fig5}. In this case direct numerical integration of the equation of motion is more efficient.

\begin{figure}[t]
\includegraphics[width=\linewidth]{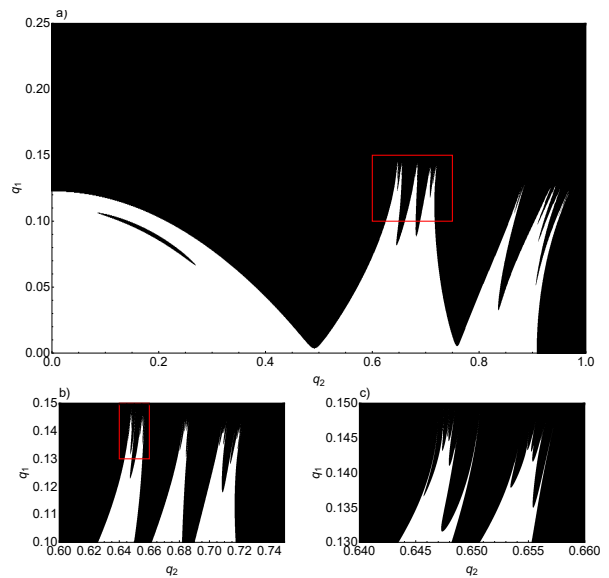}
\caption{\label{fig7} Stability diagrams for $\eta=e$, an irrational number, without damping or coupling. In a), the entire simulated $(q_1,q_2)$ range is shown. Red boxes in a) and b) diagrams indicate regions plotted in the b) and c) diagrams, respectively. The diagram in b) magnifies the region of $q_2\in[.60, .75],~q_1\in[.10, .15]$ and the diagram in c) magnifies the region of $q_2\in[.64, .66],~q_1\in[.130, .150]$. The diagram's visual complexity appears constant over each zoom step which may be due to the irrationality of $e$ and thus may continue infinitely.}
\end{figure}

The method we chose for this purpose is numerical integration of Eq.~\eqref{eq2} in \textit{Mathematica}. The numerical method then not only makes solutions with an irrational $\eta$ possible, but also allows for modifications of the equation of motion to incorporate effects such as damping, multiple interacting particles, real trap geometries, or electrical noise that heats the particles.

\begin{figure*}[t]
\includegraphics[width=\textwidth]{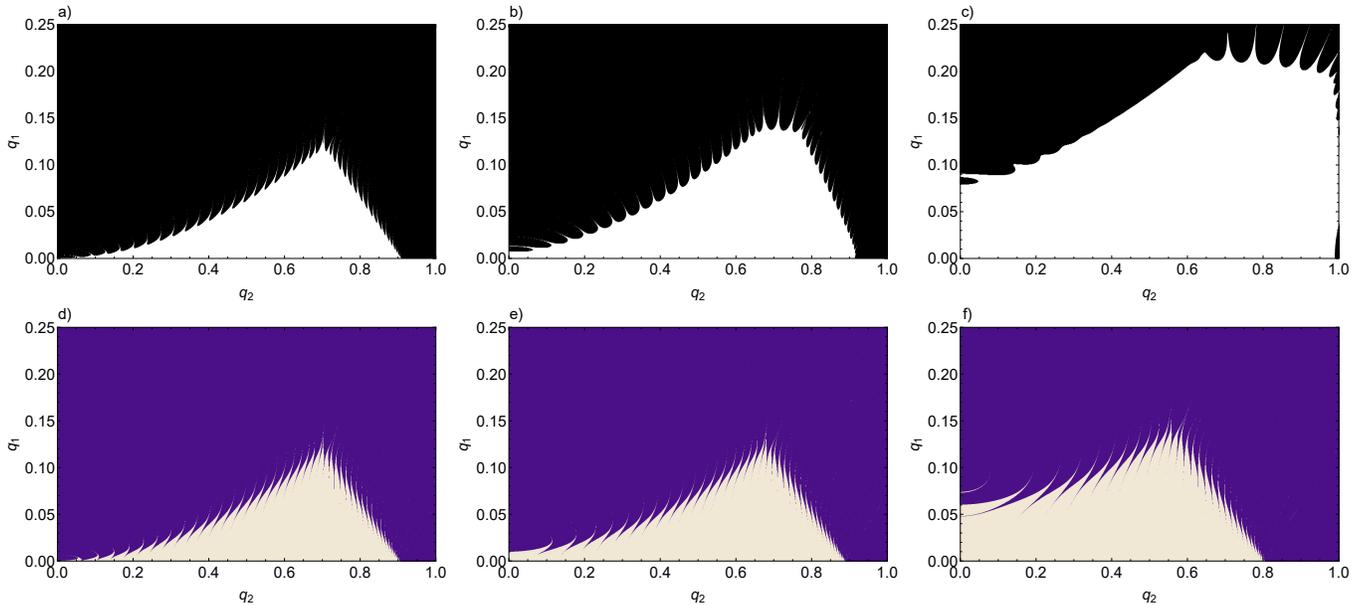}
\caption{\label{fig8}(top row) Stability diagrams with damping $b$ for $\eta=45$, evaluated with numeric integration. They correspond to $b=0$ (a)), $b=0.1$ (b)) and $b=0.4$ (c)). The stability region broadens with increased damping in such a way that the area of stability between two unstable resonances extends, widening the stable $q_2$ region and the maximum of stable $q_1$. \newline
(bottom row) Stability diagrams of a charged particle in a two-frequency Paul trap with $\eta = 45$ and an additional magnetic field $B_0$ along the z-axis (see text), calculated with the matrix determinant method. Values of the magnetic field parameter correspond to $p=0.1$ (d)), $p=0.3$ (e)) and $p=0.7$ (f)). The stability region grows larger with increased magnetic field coupling $p$ so that at $q_2=0$ the maximum value of $q_1$ increases, while the maximum value of stability for $q_2$ decreases.}
\end{figure*}

We evaluate the equation of motion for a time interval $[\tau_0,\tau_1]$, during which the amplitudes of a particle's oscillation $A_0$ at $\tau_0$ and $A_1$ at $\tau_1$ are either of the same order of magnitude or escalate to a difference of many orders of magnitude for, respectively, stable or unstable $(q_1,q_2)$ combinations. This time interval's length has to be chosen in respect to the available computation power. Simultaneously the precision with which we can determine stability increases with the length of this time interval due to the fact that a solution close to the border of stability takes a long time to diverge. We compromised between a short solution time and precision by choosing $\Delta \tau = \tau_1-\tau_0 < 1000$ (with units $2/\Omega_2$). The computer had 64 GB RAM and 10-core processor, running simulations for a time interval of $[0,600]$. 

We define the parameter $s(\tau_0,\tau_1)$ to determine whether a particular combination of $(q_1, q_2)$ is stable. This stability parameter is evaluated by integrating the square of the solution over $[\tau_0,\tau_1]$ and dividing it by the intervals length $\Delta \tau$:
\begin{equation}\label{eq14}
s(\tau_0,\tau_1) = \frac{1}{\Delta \tau} \int_{\tau_0}^{\tau_1} x(\tau)^2 d\tau.
\end{equation}
We choose $s(\tau_0,\tau_1)<10000$ to identify a combination of $(q_1,q_2)$ as stable. This condition catches oscillations that are close to the parametric resonances and have large amplitudes. For $s(\tau_0,\tau_1)>10000$ we expect the amplitude to continue growing for times beyond the integration limit $\tau_1$, but ultimately the value of this condition is arbitrary and chosen based on available computing resources. This stability parameter $s(\tau_0,\tau_1)$ is evaluated by numerically integrating the coupled equation $s'(\tau)=2x(\tau) / \Delta \tau$ simultaneously with the equation of motion. This significantly reduces computational overhead.  Using these rules to decide if a particular combination of $(q_1,q_2)$ is stable or not, we can modify the original equation of motion and consider extensions to the stability question. 

\subsection{Numeric integration vs. matrix determinant}
Without damping and coupling the numerical and matrix determinant methods agree, but the ability to treat other cases --- in general modifications to Eq.~\eqref{eq2} --- and the fact that we can use irrational frequency ratios $\eta$ is what sets the numerical method apart from the matrix determinant method. A downside of these numerical calculations for differential equations is the larger amount of time it takes to do a calculation, and the lower precision around sharp features. While the analytic solutions take around five minutes to yield results, the numerical solutions take anywhere from three to over ten hours.

The stability diagrams from numerical integration and matrix determinant calculations are compared in Fig.~\ref{fig6}. The thin resonances extending to the $q_2$-axis are less visible in the numerically calculated diagrams due to the limited $(q_1,q_2)$ resolution and finite integration time. These resonances are infinitely thin near the $q_2$-axis --- which means hardly resolvable --- and evolve over large time periods that would require excessive computation power to evaluate.

Stability diagrams for irrational values of $\eta$ share general features with diagrams for nearby integer values of $\eta$, but contain complex structures of instability that appear to have a fractal nature of scale-invariant complexity as shown in Fig.~\ref{fig7}.



\subsection{Equation of damped motion}
Adding a damping term to Eq.~\eqref{eq2} yields a new differential equation that can now model effects such as laser cooling or coupling of the particles' mechanical motions to a cold resonant circuit~\cite{Maero2012,Ulmer2013},
\begin{equation}\label{eq15}
\ddot{x}(\tau) + 2b \dot{x}(\tau) + (a - 2 q_1 \cos{2\eta^{-1} \tau} - 2 q_2 \cos{2 \tau})x(\tau)=0,
\end{equation}
where $b= \beta/m\Omega_2$ and $\beta$ is the damping parameter. Equation~\eqref{eq15} is numerically integrated and the stability assessed by the same threshold as in Eq.~\eqref{eq14}. The effect of damping terms is pictured in Fig.~\ref{fig8}. 

\begin{figure*}[t]
 \includegraphics[width=\textwidth]{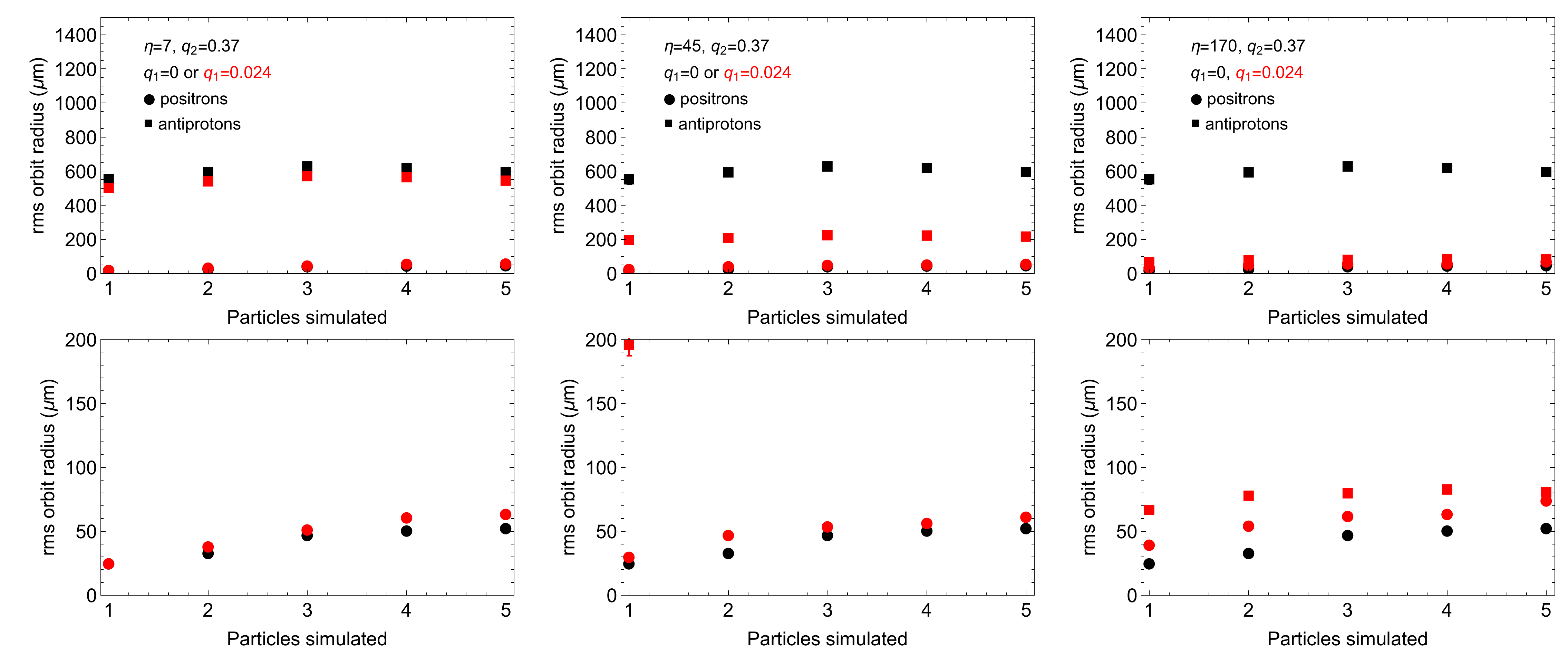}
 \caption{\label{fig9} Numerically calculated rms radii for various numbers of positrons (circle) and antiprotons (squares). 
The calculation is carried out for the same number of positrons and antiprotons inside the trap, up to five particles of each kind and with Coulomb interaction.
Ions were assumed to have an average kinetic energy corresponding to a temperature of $4$~K. Adding a low-frequency potential can significantly affect the cloud size of antiprotons, while leaving the positron cloud largely unaffected. Black symbols show rms radii for positrons and antiprotons without the low-frequency potential. Calculations were performed for $\eta = 7, 45$ and $170$. 
The bottom row shows the same data on an enlarged vertical scale. 
These plots show that for a frequency ratio $\eta = 170$ it is possible to almost match positron and antiproton cloud sizes (see Fig.~\ref{fig10}).}
\end{figure*}

\subsection{Equations of motion with magnetic field}
For antihydrogen production the ion trap must be operated in the presence of a magnetic field to trap the resulting neutral particles. A magnetic trap uses an inhomogeneous magnetic field to create a potential well. As a first approximation of the effect of the magnetic field on the charged particles, we consider the effect of a uniform magnetic field $B_0$ along the z-axis of the trap. Due to the Lorentz force the x- and y-motion are no longer independent and we get a pair of coupled equations,
\begin{equation}\label{eq16}
\ddot{x}(\tau) - p \dot{y}(\tau)+ (a - 2 q_1 \cos{2 \eta^{-1}\tau} - 2 q_2 \cos{2  \tau})x(\tau) = 0,
\end{equation}
\begin{equation}\label{eq17}
\ddot{y}(\tau) + p \dot{x}(\tau)+ (a - 2 q_1 \cos{2 \eta^{-1}\tau} - 2 q_2 \cos{2 \tau})y(\tau) = 0,
\end{equation}
where $p=(2 B_0 Z e)/(\Omega_2 m)$ is a dimensionless magnetic field parameter related to the cyclotron frequency. Stability can be evaluated using the matrix determinant method after making a coordinate transformation to a frame rotating with frequency $p/2$~\cite{Seriesa}. In this rotating frame the magnetic field acts as an extra DC potential. The effects of different values of $p$ are shown in the stability diagrams of Fig.~\ref{fig8}. In general the magnetic field increases stability due to the radially directed Lorentz force. This effect has been observed in single-frequency Paul traps combined with Penning traps~\cite{Walz1995}.

\section{Antihydrogen production in a two-frequency Paul trap}
\label{sec4}
As discussed previously, a benefit of a two-frequency trap for charged particles with vastly different charge-to-mass ratios is the independent control over the trapping potential for each species. In a single-frequency trap the heavy particle experiences a much weaker trapping potential, resulting in a larger volume of confinement and poor overlap with the light particle cloud. Adding a low-frequency field allows heavy charged particles to be compressed without affecting the light charged particles. This also opens the possibility of independent transport of charged species within the same volume. Using a nested electrode structure the two charged species may be initially trapped in different volumes and then merged. This is advantageous for antihydrogen production, where the current procedure is to trap antiprotons and positrons in independent potential wells and then inject the antiprotons into the positron cloud.

We ran numerical calculations of trapping that account for the inter-particle Coulomb interactions. For a collection of $N$ positrons and $N'$ antiprotons we introduce the variables $\mathbf{r_i}= (x_i,y_i,z_i)$ and $\mathbf{R_k} = (X_k,Y_k,Z_k)$ to indicate the position vector for the $i$-th positron or $k$-th antiproton, respectively. The equations of motions that we solve are:

\begin{align}\label{eq18}
\ddot{\mathbf{r}}_i + (a -  2 q_1 &\cos{2 \eta^{-1} \tau'} - 2 q_2 \cos{2 \tau'})\begin{pmatrix}x_i\\y_i\\-2z_i\end{pmatrix} =\\ 
&\Gamma \sum_{j\neq i}^N \frac{\mathbf{r_i-r_j}}{|\mathbf{r_i-r_j}|^3}-\Gamma \sum_{j'}^{N'} \frac{\mathbf{r_i-R_{j'}}}{|\mathbf{r_i-R_{j'}}|^3},\notag\\
\ddot{\mathbf{R}}_k - \frac{1}{\rho}(a -  2 q_1 &\cos{2 \eta^{-1} \tau'} - 2 q_2 \cos{2 \tau'})\begin{pmatrix}X_k\\Y_k\\-2Z_k\end{pmatrix} =\\ 
&\frac{\Gamma}{\rho} \sum_{l\neq k}^{N'} \frac{\mathbf{R_k-R_l}}{|\mathbf{R_k-R_l}|^3}-\frac{\Gamma}{\rho} \sum_{l'}^{N} \frac{\mathbf{R_k-r_{l'}}}{|\mathbf{R_k-r_{l'}}|^3},\notag
\end{align}
where the $\rho = m_p/m_e$ is the ratio of masses, and $\Gamma$ is a dimensionless Coulomb constant given by
\begin{equation}\label{eq19}
\Gamma = \frac{4 e^2}{m_e \Omega_2^2 l_0^3}.
\end{equation}
This constant is defined in cgs units ($e = 4.8\times10^{-10}$~statC), and the characteristic length scale $l_0$ is chosen to be $10^{-4}$~cm. This formulation of the equations of motion is inspired by the work in Ref.~\cite{Geyer2012}, where antihydrogen production in a single-frequency trap was considered. In that work production of transient, classically bound antihydrogen states was observed in numerical simulations for a single-frequency Paul trap optimized for positron confinement. Reference~\cite{Geyer2012} also claims a trapping mechanism related to the attraction between trapped positrons and antiprotons. While the Coulomb interaction certainly provides a significant attractive force between antiprotons and positrons in close proximity, we believe the effect to be overstated in \cite{Geyer2012} and they primarily observe the weak but still-significant confining force of an infinite-range dynamic potential on antiprotons. This is demonstrated in Fig.~\ref{fig9}, where a single-frequency Paul trap optimized for positrons provides weak confinement for antiprotons without positrons in the trap. 

In the first simulation, we consider a mixture composed equally of positrons and antiprotons. The equations are numerically integrated for 2 to 10 particles in total.
The root-mean-square (rms) radius, $\sqrt{\langle \mathbf{|r_i|^2}\rangle}$, of each particle's orbit in a simulation is evaluated and the average of these values over all particles is calculated. The simulation for each particle number is repeated 15 times with a randomly chosen set of initial velocities corresponding to a 4~K temperature. 
However, the temperature of particles inside the trap during an experiment is not necessarily 4~K, as heating mechanisms such as due to the RF drive in the presence of nonlinearities and the Coulomb interaction in comparison to the resistive cooling time constant may be significant.
The results of these simulations for positrons and antiprotons are shown in Fig.~\ref{fig9}. The values of $q_2 = 0.37$ and $q_1 = 0.024$ for the positrons were chosen as stable operating regions for all three frequency ratios and along all three axes. The trap drive frequencies are chosen such that $\Omega_2/2\pi = 600$~MHz and $\eta = 170$. Results were calculated with and without the $q_1$ term, and clearly show that for large frequency ratios the antiproton cloud can be compressed considerably without having any significant effect on the positron cloud. 

\begin{figure}[t]
\includegraphics[width=\linewidth]{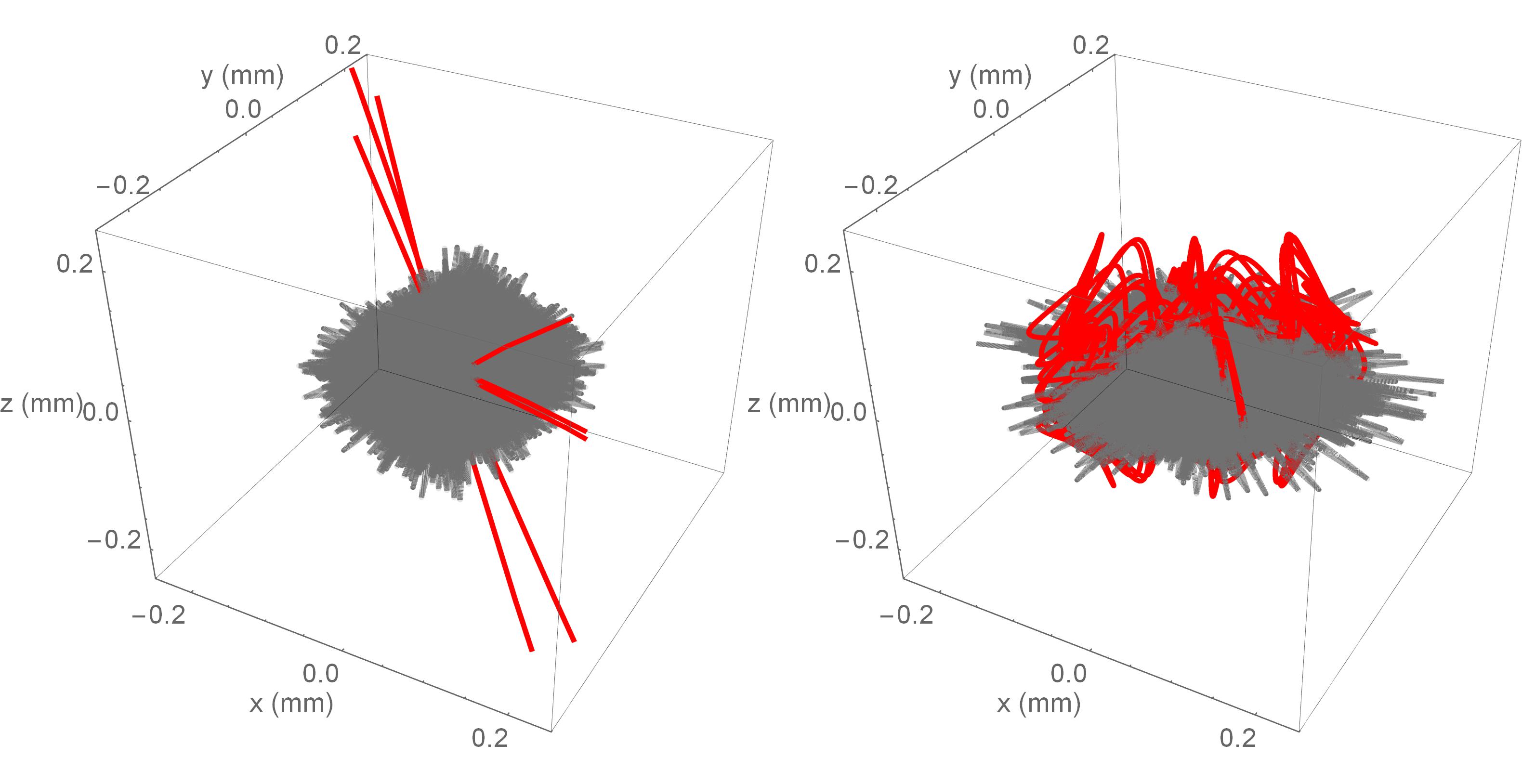}
\caption{\label{fig10} Orbits for 10 positrons (light gray) and two antiprotons (red) in a two-frequency trap with positron trap parameters $q_2 = 0.37$, $\eta = 170$, 
corresponding to $\Omega_2=2\pi \times 600~MHz$ and $\Omega_1 \approx 2\pi \times 3.53~MHz$
, $q_1 = 0$ (left) and $q_1 = 0.024$ (right). The simulation is done for a realistic scenario where both species are present. When $q_1=0$ the antiproton orbit extends outside the plot boundaries. }
\end{figure}

\begin{figure}[t]
\includegraphics[width=\linewidth]{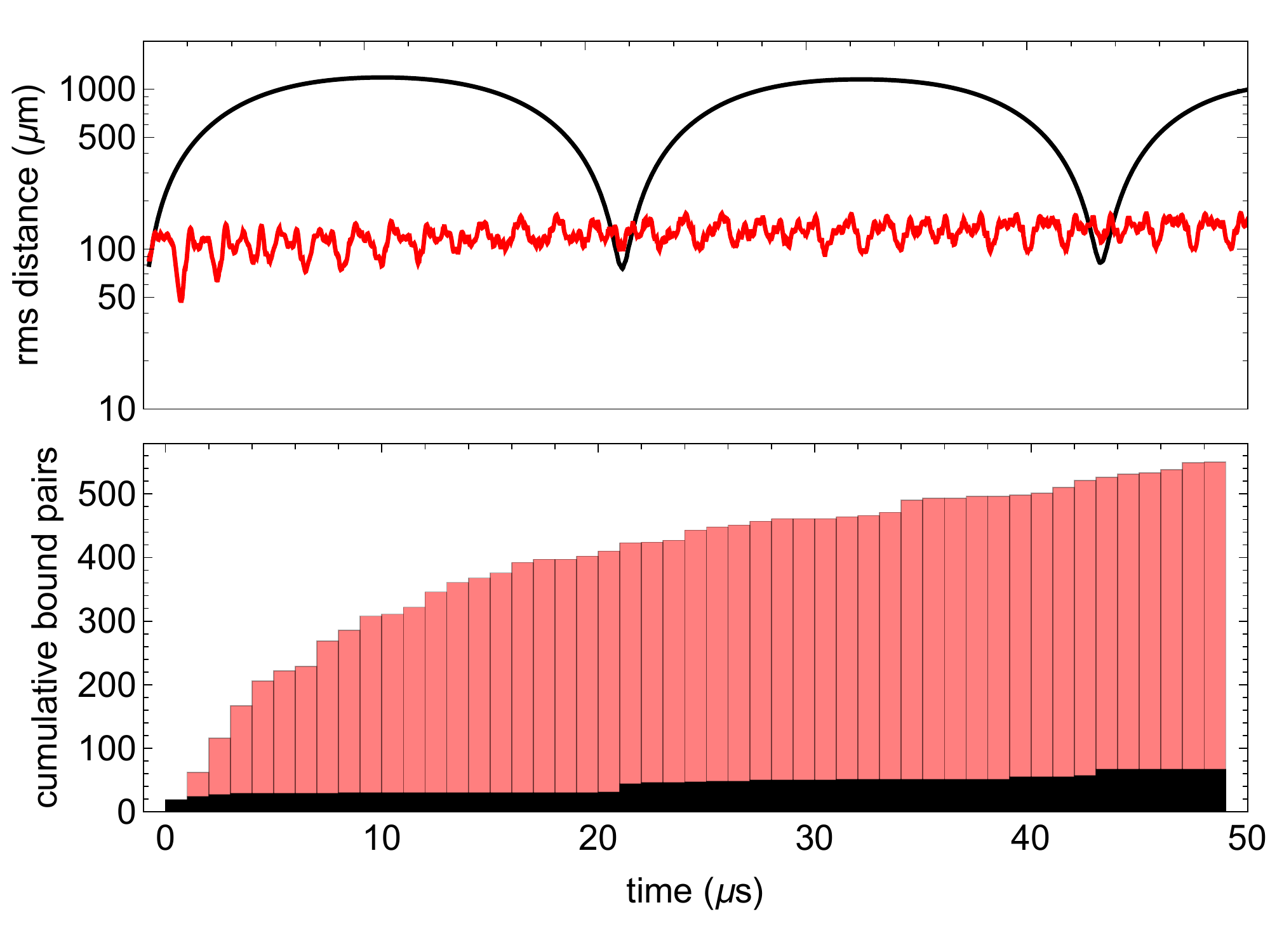}
\caption{\label{fig11} (top) Plot of average rms distance between positrons and antiprotons for positron trap parameters $q_1 = 0$ [black] and $q_1 = 0.024$ [red]. With a second AC potential applied to the trap, the maximum rms distance of an antiproton-positron pair decreases by an order of magnitude and the time period of the two particles orbiting each other is reduced by roughly a factor of 10. The frequency ratio is $\eta =  170$ and $q_2 = 0.37$. (bottom) Cumulative number of bound antiproton-positron pairs for the same trap parameters, (black) $q_1 = 0$ and (red) $q_1 = 0.024$.  The appearance of bound pairs is five times more frequent with the non-zero $q_1$.
Note that the average rms amplitude does not increase significantly, suggesting that heating on the recombination timescale can be neglected.
} 
\end{figure}

In the second simulation we consider a mixture of 10 positrons and two antiprotons using the same parameters as before. The simulation shows that recombination is likely to occur at an enhanced rate with a two-frequency trap, due primarily to the increased overlap of the antiproton and positron clouds. Figure~\ref{fig10} shows positron and antiproton orbits with and without the low-frequency potential. The low-frequency potential does seem to moderately affect the positron orbits, possibly due to the increased rate of energy changing collisions between positrons and antiprotons. The average rms distance between antiprotons and positrons is plotted in Fig.~\ref{fig11}. 

To quantify relative recombination rates, we calculate the energy of each reduced-mass positron-antiproton pair,
\begin{equation}\label{eq20}
E(t) = \frac{1}{2}\frac{m_e m_p}{m_e+m_p}(\mathbf{v_{rel}.v_{rel}}) - \frac{\Gamma}{|\mathbf{r_{rel}}|},
\end{equation}
where $\mathbf{v_{rel}}$ is the relative velocity of a given positron-antiproton pair and $\mathbf{r_{rel}}$ is the distance between them. A negative energy corresponds to a classically bound state. A list containing the energy of every positron-antiproton pair ${E_{ij}(t)}$ is calculated as a function of time, and the number of negative energies is recorded at every point in the simulation. Figure~\ref{fig11} shows the cumulative number of bound pairs produced during the simulation, with and without the low-frequency confining potential. A non-zero $q_1$ results in bound antiproton-positron pairs appearing $5\times$ more often than for $q_1 = 0$.

\subsection{Charge density}
The Coulomb interaction between positrons and antiprotons is conservative, and to create a bound state from initially unbound particles a third party must remove energy from the system. The spontaneous emission of photons is one possible mechanism, but is a slow process compared to the characteristic close-interaction time for charged particles. The primary mechanism for antihydrogen production in the ALPHA experiment is a three-body scattering process that relies on a high-density positron plasma~\cite{Gabrielse1989}. More than $10^6$ positrons are trapped with a density of $>10^7$~$\bar{e}$/cm$^3$~\cite{Amole2014}.

We estimate the achievable positron densities in a Paul trap by assuming a force balance between the Coulomb and pseudopotential forces for a positron at the edge of the positron cloud,
\begin{equation*}
\frac{N e^2}{r^2} = \frac{1}{2}m \omega^2 r,
\end{equation*}
where the equation is written in cgs units. The antiproton density is assumed negligibly low. This leads to an expected charge density $\rho_{\bar{e}} = (3/8\pi)m \omega^2/(e^2)$. If we assume a trap drive frequency of $\Omega_2/2\pi = 6\times10^8$~Hz, and $q_2 = 0.37$, the positron secular frequency is $\omega/2\pi \approx 80$~MHz and the maximum positron density is $\rho_{\bar{e}} \approx  2\times10^8$~$\bar{e}$/cm$^{3}$, significantly higher than in the ALPHA experiment. This suggests that three-body recombination is also a viable option for an all rf trap.

Extrapolating the conclusions of our simulations from several charged particles to several million charged particles is not straightforward, for instance instability may arise due to excitation of collective particle oscillations by the dynamic potential. These problems may be avoided, however, by using more compact trap configurations, for instance planar traps fabricated with atom-chip technology~\cite{Reichel2011, Keil2016}. These can localize charged particles more precisely and permit optical access without restrictions enforced by large magnets. Increased overlap and localization would facilitate studies of other recombination mechanisms, such as resonantly enhanced photoinduced recombination~\cite{Yousif1991,Wolf1993,Amoretti2006}. Recombination in a smaller volume may also simplify direct laser cooling of ground-state antihydrogen that may be produced~\cite{Kolbe2011,Michan2014}. 
Together these methods may reduce the number of positrons and antiprotons necessary, and increase the precision and rate of experiments with trapped $\bar{H}$.

\subsection{Heating, cooling and non-linear resonances in real Paul traps}

Real Paul traps have effects that perturb the quadrupole potential we considered. 
Perfect hyperbolic electrodes are hardly realizable and often approximated by spherical surfaces. Those real trap geometries make for a potential different from the perfect quadrupole and add structures of instabilities to the stability diagrams~\cite{Alheit1996a}. Mathematically, the additional instabilities can be calculated as resonances of the secular frequency with terms of the series expansion of the potential. Space charge effects of the particles being confined in a small volume of the trap also contribute to deviations from the assumed harmonic potential. The effects of non-linear resonances caused by deviations from the perfect quadrupole in Paul traps are well known for single frequencies, but have to be worked out for a two-frequency trap.

In our simulations using only up to ten particles we do not see any signatures of these instabilities, but for larger particle numbers inside the trap, the probability of collisions between particles is increased and can place the momentum of the particles out of phase with the potential. The particles then gain energy from the field instead of being forced on a stable orbit. This energy gain translates to a higher equilibrium temperature of the particle cloud and can have considerable influence on the cloud size. This rf heating effect is dependent on the particle number, ${T \propto N^{2/3}}$~\cite{ASACUSA2005}.



How big the influence of rf heating, trap imperfections and (sympathetic) cooling in our trapping constellation is, is a topic of future investigations.

\section{Summary}
We have discussed the potential of two-frequency Paul traps for the simultaneous trapping of positrons and antiprotons for recombination to antihydrogen. Stable regions in the trap parameter space have been identified and confirmed using independent methods based on Floquet theory and direct numerical integration of the equations of motion. Floquet theory provides stability maps for any rational frequency ratio, while numerical integration provides stability maps of reduced precision for any possible frequency ratio. Additional effects such as those of damping and magnetic fields were also investigated. We have further confirmed that two-frequency potentials enable charged particles with very different charge-mass ratios to be trapped simultaneously in volumes of similar size, a significant improvement over single-frequency Paul traps.  The influence of this control on the rate of antihydrogen production is a topic of continued investigation.

The feasibility of two-frequency Paul traps for antihydrogen recombination is a topic that merits further study; a number of important questions need to be answered. What effect does micromotion in an rf trap have on the energy spectrum of the produced antihydrogen? How will the trap electric fields contribute to ionization loss of Rydberg states? What trap depths can be achieved with real electrode geometries? Can atomic ion species be used for sympathetic cooling?

Investigation of recombination dynamics in two-frequency traps can be pursued initially with ions such as $^9$Be$^+$ or $^{40}$Ca$^+$ and electrons. These positive ions have convenient laser-cooling wavelengths, which simplifies many technical challenges of detection and cooling. If the techniques can eventually be proven with protons and electrons, the extension to antihydrogen is mainly a question of technical complexity and availability of antiprotons.

\begin{acknowledgments}
This work was supported in part by the DFG DIP project Ref. FO 703/2-1 1 SCHM 1049/7-1. DB, FSK, and NL acknowledge financial support by the Cluster of Excellence PRISMA at the Johannes-Gutenberg Universit\"{a}t Mainz. FSK acknowledges support from the DFG within the project BESCOOL. NL was supported by a Marie Curie International Incoming Fellowship within the 7th European Community Framework Programme.
\end{acknowledgments}
\bibliographystyle{apsrev4-1.bst}
\bibliography{Complete_Anti_H}

\begin{thebibliography}{42}%
\makeatletter
\providecommand \@ifxundefined [1]{%
 \@ifx{#1\undefined}
}%
\providecommand \@ifnum [1]{%
 \ifnum #1\expandafter \@firstoftwo
 \else \expandafter \@secondoftwo
 \fi
}%
\providecommand \@ifx [1]{%
 \ifx #1\expandafter \@firstoftwo
 \else \expandafter \@secondoftwo
 \fi
}%
\providecommand \natexlab [1]{#1}%
\providecommand \enquote  [1]{``#1''}%
\providecommand \bibnamefont  [1]{#1}%
\providecommand \bibfnamefont [1]{#1}%
\providecommand \citenamefont [1]{#1}%
\providecommand \href@noop [0]{\@secondoftwo}%
\providecommand \href [0]{\begingroup \@sanitize@url \@href}%
\providecommand \@href[1]{\@@startlink{#1}\@@href}%
\providecommand \@@href[1]{\endgroup#1\@@endlink}%
\providecommand \@sanitize@url [0]{\catcode `\\12\catcode `\$12\catcode
  `\&12\catcode `\#12\catcode `\^12\catcode `\_12\catcode `\%12\relax}%
\providecommand \@@startlink[1]{}%
\providecommand \@@endlink[0]{}%
\providecommand \url  [0]{\begingroup\@sanitize@url \@url }%
\providecommand \@url [1]{\endgroup\@href {#1}{\urlprefix }}%
\providecommand \urlprefix  [0]{URL }%
\providecommand \Eprint [0]{\href }%
\providecommand \doibase [0]{http://dx.doi.org/}%
\providecommand \selectlanguage [0]{\@gobble}%
\providecommand \bibinfo  [0]{\@secondoftwo}%
\providecommand \bibfield  [0]{\@secondoftwo}%
\providecommand \translation [1]{[#1]}%
\providecommand \BibitemOpen [0]{}%
\providecommand \bibitemStop [0]{}%
\providecommand \bibitemNoStop [0]{.\EOS\space}%
\providecommand \EOS [0]{\spacefactor3000\relax}%
\providecommand \BibitemShut  [1]{\csname bibitem#1\endcsname}%
\let\auto@bib@innerbib\@empty
\bibitem [{\citenamefont {L{\"{u}}ders}(1957)}]{Luders1957}%
  \BibitemOpen
  \bibfield  {author} {\bibinfo {author} {\bibfnamefont {G.}~\bibnamefont
  {L{\"{u}}ders}},\ }\href {\doibase 10.1016/0003-4916(57)90032-5} {\bibfield
  {journal} {\bibinfo  {journal} {Annals of Physics}\ }\textbf {\bibinfo
  {volume} {2}},\ \bibinfo {pages} {1} (\bibinfo {year} {1957})}\BibitemShut
  {NoStop}%
\bibitem [{\citenamefont {Parthey}\ \emph {et~al.}(2011)\citenamefont
  {Parthey}, \citenamefont {Matveev}, \citenamefont {Alnis}, \citenamefont
  {Bernhardt}, \citenamefont {Beyer}, \citenamefont {Holzwarth}, \citenamefont
  {Maistrou}, \citenamefont {Pohl}, \citenamefont {Predehl}, \citenamefont
  {Udem}, \citenamefont {Wilken}, \citenamefont {Kolachevsky}, \citenamefont
  {Abgrall}, \citenamefont {Rovera}, \citenamefont {Salomon}, \citenamefont
  {Laurent},\ and\ \citenamefont {H{\"{a}}nsch}}]{Parthey2011}%
  \BibitemOpen
  \bibfield  {author} {\bibinfo {author} {\bibfnamefont {C.~G.}\ \bibnamefont
  {Parthey}}, \bibinfo {author} {\bibfnamefont {A.}~\bibnamefont {Matveev}},
  \bibinfo {author} {\bibfnamefont {J.}~\bibnamefont {Alnis}}, \bibinfo
  {author} {\bibfnamefont {B.}~\bibnamefont {Bernhardt}}, \bibinfo {author}
  {\bibfnamefont {A.}~\bibnamefont {Beyer}}, \bibinfo {author} {\bibfnamefont
  {R.}~\bibnamefont {Holzwarth}}, \bibinfo {author} {\bibfnamefont
  {A.}~\bibnamefont {Maistrou}}, \bibinfo {author} {\bibfnamefont
  {R.}~\bibnamefont {Pohl}}, \bibinfo {author} {\bibfnamefont {K.}~\bibnamefont
  {Predehl}}, \bibinfo {author} {\bibfnamefont {T.}~\bibnamefont {Udem}},
  \bibinfo {author} {\bibfnamefont {T.}~\bibnamefont {Wilken}}, \bibinfo
  {author} {\bibfnamefont {N.}~\bibnamefont {Kolachevsky}}, \bibinfo {author}
  {\bibfnamefont {M.}~\bibnamefont {Abgrall}}, \bibinfo {author} {\bibfnamefont
  {D.}~\bibnamefont {Rovera}}, \bibinfo {author} {\bibfnamefont
  {C.}~\bibnamefont {Salomon}}, \bibinfo {author} {\bibfnamefont
  {P.}~\bibnamefont {Laurent}}, \ and\ \bibinfo {author} {\bibfnamefont
  {T.~W.}\ \bibnamefont {H{\"{a}}nsch}},\ }\href {\doibase
  10.1103/PhysRevLett.107.203001} {\bibfield  {journal} {\bibinfo  {journal}
  {Physical Review Letters}\ }\textbf {\bibinfo {volume} {107}},\ \bibinfo
  {pages} {1} (\bibinfo {year} {2011})},\ \Eprint
  {http://arxiv.org/abs/arXiv:1107.3101v1} {arXiv:arXiv:1107.3101v1}
  \BibitemShut {NoStop}%
\bibitem [{\citenamefont {Hardy}\ \emph {et~al.}(1979)\citenamefont {Hardy},
  \citenamefont {Berlinsky},\ and\ \citenamefont {Whitehead}}]{Hardy1979}%
  \BibitemOpen
  \bibfield  {author} {\bibinfo {author} {\bibfnamefont {W.~N.}\ \bibnamefont
  {Hardy}}, \bibinfo {author} {\bibfnamefont {A.~J.}\ \bibnamefont
  {Berlinsky}}, \ and\ \bibinfo {author} {\bibfnamefont {L.~A.}\ \bibnamefont
  {Whitehead}},\ }\href {\doibase 10.1103/PhysRevLett.42.1042} {\bibfield
  {journal} {\bibinfo  {journal} {Physical Review Letters}\ }\textbf {\bibinfo
  {volume} {42}},\ \bibinfo {pages} {1042} (\bibinfo {year}
  {1979})}\BibitemShut {NoStop}%
\bibitem [{\citenamefont {Amole}\ \emph {et~al.}(2012)\citenamefont {Amole},
  \citenamefont {Ashkezari}, \citenamefont {Baquero-Ruiz}, \citenamefont
  {Bertsche}, \citenamefont {Bowe}, \citenamefont {Butler}, \citenamefont
  {Capra}, \citenamefont {Cesar}, \citenamefont {Charlton}, \citenamefont
  {Deller}, \citenamefont {Donnan}, \citenamefont {Eriksson}, \citenamefont
  {Fajans}, \citenamefont {Friesen}, \citenamefont {Fujiwara}, \citenamefont
  {Gill}, \citenamefont {Gutierrez}, \citenamefont {Hangst}, \citenamefont
  {Hardy}, \citenamefont {Hayden}, \citenamefont {Humphries}, \citenamefont
  {Isaac}, \citenamefont {Jonsell}, \citenamefont {Kurchaninov}, \citenamefont
  {Little}, \citenamefont {Madsen}, \citenamefont {McKenna}, \citenamefont
  {Menary}, \citenamefont {Napoli}, \citenamefont {Nolan}, \citenamefont
  {Olchanski}, \citenamefont {Olin}, \citenamefont {Pusa}, \citenamefont
  {Rasmussen}, \citenamefont {Robicheaux}, \citenamefont {Sarid}, \citenamefont
  {Shields}, \citenamefont {Silveira}, \citenamefont {Stracka}, \citenamefont
  {So}, \citenamefont {Thompson}, \citenamefont {van~der Werf},\ and\
  \citenamefont {Wurtele}}]{Amole2012}%
  \BibitemOpen
  \bibfield  {author} {\bibinfo {author} {\bibfnamefont {C.}~\bibnamefont
  {Amole}}, \bibinfo {author} {\bibfnamefont {M.~D.}\ \bibnamefont
  {Ashkezari}}, \bibinfo {author} {\bibfnamefont {M.}~\bibnamefont
  {Baquero-Ruiz}}, \bibinfo {author} {\bibfnamefont {W.}~\bibnamefont
  {Bertsche}}, \bibinfo {author} {\bibfnamefont {P.~D.}\ \bibnamefont {Bowe}},
  \bibinfo {author} {\bibfnamefont {E.}~\bibnamefont {Butler}}, \bibinfo
  {author} {\bibfnamefont {a.}~\bibnamefont {Capra}}, \bibinfo {author}
  {\bibfnamefont {C.~L.}\ \bibnamefont {Cesar}}, \bibinfo {author}
  {\bibfnamefont {M.}~\bibnamefont {Charlton}}, \bibinfo {author}
  {\bibfnamefont {a.}~\bibnamefont {Deller}}, \bibinfo {author} {\bibfnamefont
  {P.~H.}\ \bibnamefont {Donnan}}, \bibinfo {author} {\bibfnamefont
  {S.}~\bibnamefont {Eriksson}}, \bibinfo {author} {\bibfnamefont
  {J.}~\bibnamefont {Fajans}}, \bibinfo {author} {\bibfnamefont
  {T.}~\bibnamefont {Friesen}}, \bibinfo {author} {\bibfnamefont {M.~C.}\
  \bibnamefont {Fujiwara}}, \bibinfo {author} {\bibfnamefont {D.~R.}\
  \bibnamefont {Gill}}, \bibinfo {author} {\bibfnamefont {a.}~\bibnamefont
  {Gutierrez}}, \bibinfo {author} {\bibfnamefont {J.~S.}\ \bibnamefont
  {Hangst}}, \bibinfo {author} {\bibfnamefont {W.~N.}\ \bibnamefont {Hardy}},
  \bibinfo {author} {\bibfnamefont {M.~E.}\ \bibnamefont {Hayden}}, \bibinfo
  {author} {\bibfnamefont {a.~J.}\ \bibnamefont {Humphries}}, \bibinfo {author}
  {\bibfnamefont {C.~a.}\ \bibnamefont {Isaac}}, \bibinfo {author}
  {\bibfnamefont {S.}~\bibnamefont {Jonsell}}, \bibinfo {author} {\bibfnamefont
  {L.}~\bibnamefont {Kurchaninov}}, \bibinfo {author} {\bibfnamefont
  {a.}~\bibnamefont {Little}}, \bibinfo {author} {\bibfnamefont
  {N.}~\bibnamefont {Madsen}}, \bibinfo {author} {\bibfnamefont {J.~T.~K.}\
  \bibnamefont {McKenna}}, \bibinfo {author} {\bibfnamefont {S.}~\bibnamefont
  {Menary}}, \bibinfo {author} {\bibfnamefont {S.~C.}\ \bibnamefont {Napoli}},
  \bibinfo {author} {\bibfnamefont {P.}~\bibnamefont {Nolan}}, \bibinfo
  {author} {\bibfnamefont {K.}~\bibnamefont {Olchanski}}, \bibinfo {author}
  {\bibfnamefont {a.}~\bibnamefont {Olin}}, \bibinfo {author} {\bibfnamefont
  {P.}~\bibnamefont {Pusa}}, \bibinfo {author} {\bibfnamefont {C.~{\O}.}\
  \bibnamefont {Rasmussen}}, \bibinfo {author} {\bibfnamefont {F.}~\bibnamefont
  {Robicheaux}}, \bibinfo {author} {\bibfnamefont {E.}~\bibnamefont {Sarid}},
  \bibinfo {author} {\bibfnamefont {C.~R.}\ \bibnamefont {Shields}}, \bibinfo
  {author} {\bibfnamefont {D.~M.}\ \bibnamefont {Silveira}}, \bibinfo {author}
  {\bibfnamefont {S.}~\bibnamefont {Stracka}}, \bibinfo {author} {\bibfnamefont
  {C.}~\bibnamefont {So}}, \bibinfo {author} {\bibfnamefont {R.~I.}\
  \bibnamefont {Thompson}}, \bibinfo {author} {\bibfnamefont {D.~P.}\
  \bibnamefont {van~der Werf}}, \ and\ \bibinfo {author} {\bibfnamefont
  {J.~S.}\ \bibnamefont {Wurtele}},\ }\href {\doibase 10.1038/nature10942}
  {\bibfield  {journal} {\bibinfo  {journal} {Nature}\ }\textbf {\bibinfo
  {volume} {483}},\ \bibinfo {pages} {439} (\bibinfo {year}
  {2012})}\BibitemShut {NoStop}%
\bibitem [{\citenamefont {Andresen}\ \emph {et~al.}(2010)\citenamefont
  {Andresen}, \citenamefont {Ashkezari}, \citenamefont {Baquero-Ruiz},
  \citenamefont {Bertsche}, \citenamefont {Bowe}, \citenamefont {Butler},
  \citenamefont {Cesar}, \citenamefont {Chapman}, \citenamefont {Charlton},
  \citenamefont {Deller}, \citenamefont {Eriksson}, \citenamefont {Fajans},
  \citenamefont {Friesen}, \citenamefont {Fujiwara}, \citenamefont {Gill},
  \citenamefont {Gutierrez}, \citenamefont {Hangst}, \citenamefont {Hardy},
  \citenamefont {Hayden}, \citenamefont {Humphries}, \citenamefont {Hydomako},
  \citenamefont {Jenkins}, \citenamefont {Jonsell}, \citenamefont
  {J{\o}rgensen}, \citenamefont {Kurchaninov}, \citenamefont {Madsen},
  \citenamefont {Menary}, \citenamefont {Nolan}, \citenamefont {Olchanski},
  \citenamefont {Olin}, \citenamefont {Povilus}, \citenamefont {Pusa},
  \citenamefont {Robicheaux}, \citenamefont {Sarid}, \citenamefont {{Seif el
  Nasr}}, \citenamefont {Silveira}, \citenamefont {So}, \citenamefont {Storey},
  \citenamefont {Thompson}, \citenamefont {van~der Werf}, \citenamefont
  {Wurtele},\ and\ \citenamefont {Yamazaki}}]{Andresen2010}%
  \BibitemOpen
  \bibfield  {author} {\bibinfo {author} {\bibfnamefont {G.~B.}\ \bibnamefont
  {Andresen}}, \bibinfo {author} {\bibfnamefont {M.~D.}\ \bibnamefont
  {Ashkezari}}, \bibinfo {author} {\bibfnamefont {M.}~\bibnamefont
  {Baquero-Ruiz}}, \bibinfo {author} {\bibfnamefont {W.}~\bibnamefont
  {Bertsche}}, \bibinfo {author} {\bibfnamefont {P.~D.}\ \bibnamefont {Bowe}},
  \bibinfo {author} {\bibfnamefont {E.}~\bibnamefont {Butler}}, \bibinfo
  {author} {\bibfnamefont {C.~L.}\ \bibnamefont {Cesar}}, \bibinfo {author}
  {\bibfnamefont {S.}~\bibnamefont {Chapman}}, \bibinfo {author} {\bibfnamefont
  {M.}~\bibnamefont {Charlton}}, \bibinfo {author} {\bibfnamefont
  {A.}~\bibnamefont {Deller}}, \bibinfo {author} {\bibfnamefont
  {S.}~\bibnamefont {Eriksson}}, \bibinfo {author} {\bibfnamefont
  {J.}~\bibnamefont {Fajans}}, \bibinfo {author} {\bibfnamefont
  {T.}~\bibnamefont {Friesen}}, \bibinfo {author} {\bibfnamefont {M.~C.}\
  \bibnamefont {Fujiwara}}, \bibinfo {author} {\bibfnamefont {D.~R.}\
  \bibnamefont {Gill}}, \bibinfo {author} {\bibfnamefont {A.}~\bibnamefont
  {Gutierrez}}, \bibinfo {author} {\bibfnamefont {J.~S.}\ \bibnamefont
  {Hangst}}, \bibinfo {author} {\bibfnamefont {W.~N.}\ \bibnamefont {Hardy}},
  \bibinfo {author} {\bibfnamefont {M.~E.}\ \bibnamefont {Hayden}}, \bibinfo
  {author} {\bibfnamefont {A.~J.}\ \bibnamefont {Humphries}}, \bibinfo {author}
  {\bibfnamefont {R.}~\bibnamefont {Hydomako}}, \bibinfo {author}
  {\bibfnamefont {M.~J.}\ \bibnamefont {Jenkins}}, \bibinfo {author}
  {\bibfnamefont {S.}~\bibnamefont {Jonsell}}, \bibinfo {author} {\bibfnamefont
  {L.~V.}\ \bibnamefont {J{\o}rgensen}}, \bibinfo {author} {\bibfnamefont
  {L.}~\bibnamefont {Kurchaninov}}, \bibinfo {author} {\bibfnamefont
  {N.}~\bibnamefont {Madsen}}, \bibinfo {author} {\bibfnamefont
  {S.}~\bibnamefont {Menary}}, \bibinfo {author} {\bibfnamefont
  {P.}~\bibnamefont {Nolan}}, \bibinfo {author} {\bibfnamefont
  {K.}~\bibnamefont {Olchanski}}, \bibinfo {author} {\bibfnamefont
  {A.}~\bibnamefont {Olin}}, \bibinfo {author} {\bibfnamefont {A.}~\bibnamefont
  {Povilus}}, \bibinfo {author} {\bibfnamefont {P.}~\bibnamefont {Pusa}},
  \bibinfo {author} {\bibfnamefont {F.}~\bibnamefont {Robicheaux}}, \bibinfo
  {author} {\bibfnamefont {E.}~\bibnamefont {Sarid}}, \bibinfo {author}
  {\bibfnamefont {S.}~\bibnamefont {{Seif el Nasr}}}, \bibinfo {author}
  {\bibfnamefont {D.~M.}\ \bibnamefont {Silveira}}, \bibinfo {author}
  {\bibfnamefont {C.}~\bibnamefont {So}}, \bibinfo {author} {\bibfnamefont
  {J.~W.}\ \bibnamefont {Storey}}, \bibinfo {author} {\bibfnamefont {R.~I.}\
  \bibnamefont {Thompson}}, \bibinfo {author} {\bibfnamefont {D.~P.}\
  \bibnamefont {van~der Werf}}, \bibinfo {author} {\bibfnamefont {J.~S.}\
  \bibnamefont {Wurtele}}, \ and\ \bibinfo {author} {\bibfnamefont
  {Y.}~\bibnamefont {Yamazaki}},\ }\href {\doibase 10.1038/nature09610}
  {\bibfield  {journal} {\bibinfo  {journal} {Nature}\ }\textbf {\bibinfo
  {volume} {468}},\ \bibinfo {pages} {673} (\bibinfo {year}
  {2010})}\BibitemShut {NoStop}%
\bibitem [{\citenamefont {Andresen}\ \emph {et~al.}(2011)\citenamefont
  {Andresen}, \citenamefont {Ashkezari}, \citenamefont {Baquero-Ruiz},
  \citenamefont {Bertsche}, \citenamefont {Bowe}, \citenamefont {Butler},
  \citenamefont {Cesar}, \citenamefont {Charlton}, \citenamefont {Deller},
  \citenamefont {Eriksson}, \citenamefont {Fajans}, \citenamefont {Friesen},
  \citenamefont {Fujiwara}, \citenamefont {Gill}, \citenamefont {Gutierrez},
  \citenamefont {Hangst}, \citenamefont {Hardy}, \citenamefont {Hayano},
  \citenamefont {Hayden}, \citenamefont {Humphries}, \citenamefont {Hydomako},
  \citenamefont {Jonsell}, \citenamefont {Kemp}, \citenamefont {Kurchaninov},
  \citenamefont {Madsen}, \citenamefont {Menary}, \citenamefont {Nolan},
  \citenamefont {Olchanski}, \citenamefont {Olin}, \citenamefont {Pusa},
  \citenamefont {Rasmussen}, \citenamefont {Robicheaux}, \citenamefont {Sarid},
  \citenamefont {Silveira}, \citenamefont {So}, \citenamefont {Storey},
  \citenamefont {Thompson}, \citenamefont {van~der Werf}, \citenamefont
  {Wurtele},\ and\ \citenamefont {Yamazaki}}]{Andresen2011}%
  \BibitemOpen
  \bibfield  {author} {\bibinfo {author} {\bibfnamefont {G.~B.}\ \bibnamefont
  {Andresen}}, \bibinfo {author} {\bibfnamefont {M.~D.}\ \bibnamefont
  {Ashkezari}}, \bibinfo {author} {\bibfnamefont {M.}~\bibnamefont
  {Baquero-Ruiz}}, \bibinfo {author} {\bibfnamefont {W.}~\bibnamefont
  {Bertsche}}, \bibinfo {author} {\bibfnamefont {P.~D.}\ \bibnamefont {Bowe}},
  \bibinfo {author} {\bibfnamefont {E.}~\bibnamefont {Butler}}, \bibinfo
  {author} {\bibfnamefont {C.~L.}\ \bibnamefont {Cesar}}, \bibinfo {author}
  {\bibfnamefont {M.}~\bibnamefont {Charlton}}, \bibinfo {author}
  {\bibfnamefont {A.}~\bibnamefont {Deller}}, \bibinfo {author} {\bibfnamefont
  {S.}~\bibnamefont {Eriksson}}, \bibinfo {author} {\bibfnamefont
  {J.}~\bibnamefont {Fajans}}, \bibinfo {author} {\bibfnamefont
  {T.}~\bibnamefont {Friesen}}, \bibinfo {author} {\bibfnamefont {M.~C.}\
  \bibnamefont {Fujiwara}}, \bibinfo {author} {\bibfnamefont {D.~R.}\
  \bibnamefont {Gill}}, \bibinfo {author} {\bibfnamefont {A.}~\bibnamefont
  {Gutierrez}}, \bibinfo {author} {\bibfnamefont {J.~S.}\ \bibnamefont
  {Hangst}}, \bibinfo {author} {\bibfnamefont {W.~N.}\ \bibnamefont {Hardy}},
  \bibinfo {author} {\bibfnamefont {R.~S.}\ \bibnamefont {Hayano}}, \bibinfo
  {author} {\bibfnamefont {M.~E.}\ \bibnamefont {Hayden}}, \bibinfo {author}
  {\bibfnamefont {A.~J.}\ \bibnamefont {Humphries}}, \bibinfo {author}
  {\bibfnamefont {R.}~\bibnamefont {Hydomako}}, \bibinfo {author}
  {\bibfnamefont {S.}~\bibnamefont {Jonsell}}, \bibinfo {author} {\bibfnamefont
  {S.~L.}\ \bibnamefont {Kemp}}, \bibinfo {author} {\bibfnamefont
  {L.}~\bibnamefont {Kurchaninov}}, \bibinfo {author} {\bibfnamefont
  {N.}~\bibnamefont {Madsen}}, \bibinfo {author} {\bibfnamefont
  {S.}~\bibnamefont {Menary}}, \bibinfo {author} {\bibfnamefont
  {P.}~\bibnamefont {Nolan}}, \bibinfo {author} {\bibfnamefont
  {K.}~\bibnamefont {Olchanski}}, \bibinfo {author} {\bibfnamefont
  {A.}~\bibnamefont {Olin}}, \bibinfo {author} {\bibfnamefont {P.}~\bibnamefont
  {Pusa}}, \bibinfo {author} {\bibfnamefont {C.~{\O}.}\ \bibnamefont
  {Rasmussen}}, \bibinfo {author} {\bibfnamefont {F.}~\bibnamefont
  {Robicheaux}}, \bibinfo {author} {\bibfnamefont {E.}~\bibnamefont {Sarid}},
  \bibinfo {author} {\bibfnamefont {D.~M.}\ \bibnamefont {Silveira}}, \bibinfo
  {author} {\bibfnamefont {C.}~\bibnamefont {So}}, \bibinfo {author}
  {\bibfnamefont {J.~W.}\ \bibnamefont {Storey}}, \bibinfo {author}
  {\bibfnamefont {R.~I.}\ \bibnamefont {Thompson}}, \bibinfo {author}
  {\bibfnamefont {D.~P.}\ \bibnamefont {van~der Werf}}, \bibinfo {author}
  {\bibfnamefont {J.~S.}\ \bibnamefont {Wurtele}}, \ and\ \bibinfo {author}
  {\bibfnamefont {Y.}~\bibnamefont {Yamazaki}},\ }\href {\doibase
  10.1038/nphys2025} {\bibfield  {journal} {\bibinfo  {journal} {Nature
  Physics}\ }\textbf {\bibinfo {volume} {7}},\ \bibinfo {pages} {558} (\bibinfo
  {year} {2011})}\BibitemShut {NoStop}%
\bibitem [{\citenamefont {Storry}\ \emph {et~al.}(2004)\citenamefont {Storry},
  \citenamefont {Speck}, \citenamefont {{Le Sage}}, \citenamefont {Guise},
  \citenamefont {Gabrielse}, \citenamefont {Grzonka}, \citenamefont {Oelert},
  \citenamefont {Schepers}, \citenamefont {Sefzick}, \citenamefont {Pittner},
  \citenamefont {Herrmann}, \citenamefont {Walz}, \citenamefont {H{\"{a}}nsch},
  \citenamefont {Comeau},\ and\ \citenamefont {Hessels}}]{Storry2004}%
  \BibitemOpen
  \bibfield  {author} {\bibinfo {author} {\bibfnamefont {C.~H.}\ \bibnamefont
  {Storry}}, \bibinfo {author} {\bibfnamefont {A.}~\bibnamefont {Speck}},
  \bibinfo {author} {\bibfnamefont {D.}~\bibnamefont {{Le Sage}}}, \bibinfo
  {author} {\bibfnamefont {N.}~\bibnamefont {Guise}}, \bibinfo {author}
  {\bibfnamefont {G.}~\bibnamefont {Gabrielse}}, \bibinfo {author}
  {\bibfnamefont {D.}~\bibnamefont {Grzonka}}, \bibinfo {author} {\bibfnamefont
  {W.}~\bibnamefont {Oelert}}, \bibinfo {author} {\bibfnamefont
  {G.}~\bibnamefont {Schepers}}, \bibinfo {author} {\bibfnamefont
  {T.}~\bibnamefont {Sefzick}}, \bibinfo {author} {\bibfnamefont
  {H.}~\bibnamefont {Pittner}}, \bibinfo {author} {\bibfnamefont
  {M.}~\bibnamefont {Herrmann}}, \bibinfo {author} {\bibfnamefont
  {J.}~\bibnamefont {Walz}}, \bibinfo {author} {\bibfnamefont {T.~W.}\
  \bibnamefont {H{\"{a}}nsch}}, \bibinfo {author} {\bibfnamefont
  {D.}~\bibnamefont {Comeau}}, \ and\ \bibinfo {author} {\bibfnamefont {E.~a.}\
  \bibnamefont {Hessels}},\ }\href {\doibase 10.1103/PhysRevLett.93.263401}
  {\bibfield  {journal} {\bibinfo  {journal} {Physical Review Letters}\
  }\textbf {\bibinfo {volume} {93}},\ \bibinfo {pages} {1} (\bibinfo {year}
  {2004})}\BibitemShut {NoStop}%
\bibitem [{\citenamefont {Gabrielse}\ \emph {et~al.}(2012)\citenamefont
  {Gabrielse}, \citenamefont {Kalra}, \citenamefont {Kolthammer}, \citenamefont
  {McConnell}, \citenamefont {Richerme}, \citenamefont {Grzonka}, \citenamefont
  {Oelert}, \citenamefont {Sefzick}, \citenamefont {Zielinski}, \citenamefont
  {Fitzakerley}, \citenamefont {George}, \citenamefont {Hessels}, \citenamefont
  {Storry}, \citenamefont {Weel}, \citenamefont {M{\"{u}}llers},\ and\
  \citenamefont {Walz}}]{Gabrielse2012}%
  \BibitemOpen
  \bibfield  {author} {\bibinfo {author} {\bibfnamefont {G.}~\bibnamefont
  {Gabrielse}}, \bibinfo {author} {\bibfnamefont {R.}~\bibnamefont {Kalra}},
  \bibinfo {author} {\bibfnamefont {W.~S.}\ \bibnamefont {Kolthammer}},
  \bibinfo {author} {\bibfnamefont {R.}~\bibnamefont {McConnell}}, \bibinfo
  {author} {\bibfnamefont {P.}~\bibnamefont {Richerme}}, \bibinfo {author}
  {\bibfnamefont {D.}~\bibnamefont {Grzonka}}, \bibinfo {author} {\bibfnamefont
  {W.}~\bibnamefont {Oelert}}, \bibinfo {author} {\bibfnamefont
  {T.}~\bibnamefont {Sefzick}}, \bibinfo {author} {\bibfnamefont
  {M.}~\bibnamefont {Zielinski}}, \bibinfo {author} {\bibfnamefont {D.~W.}\
  \bibnamefont {Fitzakerley}}, \bibinfo {author} {\bibfnamefont {M.~C.}\
  \bibnamefont {George}}, \bibinfo {author} {\bibfnamefont {E.~a.}\
  \bibnamefont {Hessels}}, \bibinfo {author} {\bibfnamefont {C.~H.}\
  \bibnamefont {Storry}}, \bibinfo {author} {\bibfnamefont {M.}~\bibnamefont
  {Weel}}, \bibinfo {author} {\bibfnamefont {a.}~\bibnamefont {M{\"{u}}llers}},
  \ and\ \bibinfo {author} {\bibfnamefont {J.}~\bibnamefont {Walz}},\ }\href
  {\doibase 10.1103/PhysRevLett.108.113002} {\bibfield  {journal} {\bibinfo
  {journal} {Physical Review Letters}\ }\textbf {\bibinfo {volume} {108}},\
  \bibinfo {pages} {12} (\bibinfo {year} {2012})},\ \Eprint
  {http://arxiv.org/abs/1201.2717} {arXiv:1201.2717} \BibitemShut {NoStop}%
\bibitem [{\citenamefont {Amole}\ \emph {et~al.}(2013)\citenamefont {Amole},
  \citenamefont {Ashkezari}, \citenamefont {Baquero-Ruiz}, \citenamefont
  {Bertsche}, \citenamefont {Butler}, \citenamefont {Capra}, \citenamefont
  {Cesar}, \citenamefont {Charlton}, \citenamefont {Deller}, \citenamefont
  {Eriksson}, \citenamefont {Fajans}, \citenamefont {Friesen}, \citenamefont
  {Fujiwara}, \citenamefont {Gill}, \citenamefont {Gutierrez}, \citenamefont
  {Hangst}, \citenamefont {Hardy}, \citenamefont {Hayden}, \citenamefont
  {Isaac}, \citenamefont {Jonsell}, \citenamefont {Kurchaninov}, \citenamefont
  {Little}, \citenamefont {Madsen}, \citenamefont {McKenna}, \citenamefont
  {Menary}, \citenamefont {Napoli}, \citenamefont {Olchanski}, \citenamefont
  {Olin}, \citenamefont {Pusa}, \citenamefont {Rasmussen}, \citenamefont
  {Robicheaux}, \citenamefont {Sarid}, \citenamefont {Shields}, \citenamefont
  {Silveira}, \citenamefont {So}, \citenamefont {Stracka}, \citenamefont
  {Thompson}, \citenamefont {van~der Werf}, \citenamefont {Wurtele},
  \citenamefont {Zhmoginov}, \citenamefont {Friedland},\ and\ \citenamefont
  {Collaboration)}}]{Amole2013}%
  \BibitemOpen
  \bibfield  {author} {\bibinfo {author} {\bibfnamefont {C.}~\bibnamefont
  {Amole}}, \bibinfo {author} {\bibfnamefont {M.~D.}\ \bibnamefont
  {Ashkezari}}, \bibinfo {author} {\bibfnamefont {M.}~\bibnamefont
  {Baquero-Ruiz}}, \bibinfo {author} {\bibfnamefont {W.}~\bibnamefont
  {Bertsche}}, \bibinfo {author} {\bibfnamefont {E.}~\bibnamefont {Butler}},
  \bibinfo {author} {\bibfnamefont {A.}~\bibnamefont {Capra}}, \bibinfo
  {author} {\bibfnamefont {C.~L.}\ \bibnamefont {Cesar}}, \bibinfo {author}
  {\bibfnamefont {M.}~\bibnamefont {Charlton}}, \bibinfo {author}
  {\bibfnamefont {A.}~\bibnamefont {Deller}}, \bibinfo {author} {\bibfnamefont
  {S.}~\bibnamefont {Eriksson}}, \bibinfo {author} {\bibfnamefont
  {J.}~\bibnamefont {Fajans}}, \bibinfo {author} {\bibfnamefont
  {T.}~\bibnamefont {Friesen}}, \bibinfo {author} {\bibfnamefont {M.~C.}\
  \bibnamefont {Fujiwara}}, \bibinfo {author} {\bibfnamefont {D.~R.}\
  \bibnamefont {Gill}}, \bibinfo {author} {\bibfnamefont {A.}~\bibnamefont
  {Gutierrez}}, \bibinfo {author} {\bibfnamefont {J.~S.}\ \bibnamefont
  {Hangst}}, \bibinfo {author} {\bibfnamefont {W.~N.}\ \bibnamefont {Hardy}},
  \bibinfo {author} {\bibfnamefont {M.~E.}\ \bibnamefont {Hayden}}, \bibinfo
  {author} {\bibfnamefont {C.~A.}\ \bibnamefont {Isaac}}, \bibinfo {author}
  {\bibfnamefont {S.}~\bibnamefont {Jonsell}}, \bibinfo {author} {\bibfnamefont
  {L.}~\bibnamefont {Kurchaninov}}, \bibinfo {author} {\bibfnamefont
  {A.}~\bibnamefont {Little}}, \bibinfo {author} {\bibfnamefont
  {N.}~\bibnamefont {Madsen}}, \bibinfo {author} {\bibfnamefont {J.~T.~K.}\
  \bibnamefont {McKenna}}, \bibinfo {author} {\bibfnamefont {S.}~\bibnamefont
  {Menary}}, \bibinfo {author} {\bibfnamefont {S.~C.}\ \bibnamefont {Napoli}},
  \bibinfo {author} {\bibfnamefont {K.}~\bibnamefont {Olchanski}}, \bibinfo
  {author} {\bibfnamefont {A.}~\bibnamefont {Olin}}, \bibinfo {author}
  {\bibfnamefont {P.}~\bibnamefont {Pusa}}, \bibinfo {author} {\bibfnamefont
  {C.~{\O}.}\ \bibnamefont {Rasmussen}}, \bibinfo {author} {\bibfnamefont
  {F.}~\bibnamefont {Robicheaux}}, \bibinfo {author} {\bibfnamefont
  {E.}~\bibnamefont {Sarid}}, \bibinfo {author} {\bibfnamefont {C.~R.}\
  \bibnamefont {Shields}}, \bibinfo {author} {\bibfnamefont {D.~M.}\
  \bibnamefont {Silveira}}, \bibinfo {author} {\bibfnamefont {C.}~\bibnamefont
  {So}}, \bibinfo {author} {\bibfnamefont {S.}~\bibnamefont {Stracka}},
  \bibinfo {author} {\bibfnamefont {R.~I.}\ \bibnamefont {Thompson}}, \bibinfo
  {author} {\bibfnamefont {D.~P.}\ \bibnamefont {van~der Werf}}, \bibinfo
  {author} {\bibfnamefont {J.~S.}\ \bibnamefont {Wurtele}}, \bibinfo {author}
  {\bibfnamefont {A.~I.}\ \bibnamefont {Zhmoginov}}, \bibinfo {author}
  {\bibfnamefont {L.}~\bibnamefont {Friedland}}, \ and\ \bibinfo {author}
  {\bibfnamefont {A.}~\bibnamefont {Collaboration)}},\ }\href {\doibase
  10.1063/1.4801067} {\bibfield  {journal} {\bibinfo  {journal} {Physics of
  Plasmas}\ }\textbf {\bibinfo {volume} {20}},\ \bibinfo {pages} {43510}
  (\bibinfo {year} {2013})}\BibitemShut {NoStop}%
\bibitem [{\citenamefont {Amole}\ \emph
  {et~al.}(2014{\natexlab{a}})\citenamefont {Amole}, \citenamefont {Andresen},
  \citenamefont {Ashkezari}, \citenamefont {Baquero-Ruiz}, \citenamefont
  {Bertsche}, \citenamefont {Bowe}, \citenamefont {Butler}, \citenamefont
  {Capra}, \citenamefont {Carpenter}, \citenamefont {Cesar}, \citenamefont
  {Chapman}, \citenamefont {Charlton}, \citenamefont {Deller}, \citenamefont
  {Eriksson}, \citenamefont {Escallier}, \citenamefont {Fajans}, \citenamefont
  {Friesen}, \citenamefont {Fujiwara}, \citenamefont {Gill}, \citenamefont
  {Gutierrez}, \citenamefont {Hangst}, \citenamefont {Hardy}, \citenamefont
  {Hayano}, \citenamefont {Hayden}, \citenamefont {Humphries}, \citenamefont
  {Hurt}, \citenamefont {Hydomako}, \citenamefont {Isaac}, \citenamefont
  {Jenkins}, \citenamefont {Jonsell}, \citenamefont {J??rgensen}, \citenamefont
  {Kerrigan}, \citenamefont {Kurchaninov}, \citenamefont {Madsen},
  \citenamefont {Marone}, \citenamefont {McKenna}, \citenamefont {Menary},
  \citenamefont {Nolan}, \citenamefont {Olchanski}, \citenamefont {Olin},
  \citenamefont {Parker}, \citenamefont {Povilus}, \citenamefont {Pusa},
  \citenamefont {Robicheaux}, \citenamefont {Sarid}, \citenamefont {Seddon},
  \citenamefont {{Seif El Nasr}}, \citenamefont {Silveira}, \citenamefont {So},
  \citenamefont {Storey}, \citenamefont {Thompson}, \citenamefont {Thornhill},
  \citenamefont {Wells}, \citenamefont {van~der Werf}, \citenamefont
  {Wurtele},\ and\ \citenamefont {Yamazaki}}]{Amole2014}%
  \BibitemOpen
  \bibfield  {author} {\bibinfo {author} {\bibfnamefont {C.}~\bibnamefont
  {Amole}}, \bibinfo {author} {\bibfnamefont {G.~B.}\ \bibnamefont {Andresen}},
  \bibinfo {author} {\bibfnamefont {M.~D.}\ \bibnamefont {Ashkezari}}, \bibinfo
  {author} {\bibfnamefont {M.}~\bibnamefont {Baquero-Ruiz}}, \bibinfo {author}
  {\bibfnamefont {W.}~\bibnamefont {Bertsche}}, \bibinfo {author}
  {\bibfnamefont {P.~D.}\ \bibnamefont {Bowe}}, \bibinfo {author}
  {\bibfnamefont {E.}~\bibnamefont {Butler}}, \bibinfo {author} {\bibfnamefont
  {A.}~\bibnamefont {Capra}}, \bibinfo {author} {\bibfnamefont {P.~T.}\
  \bibnamefont {Carpenter}}, \bibinfo {author} {\bibfnamefont {C.~L.}\
  \bibnamefont {Cesar}}, \bibinfo {author} {\bibfnamefont {S.}~\bibnamefont
  {Chapman}}, \bibinfo {author} {\bibfnamefont {M.}~\bibnamefont {Charlton}},
  \bibinfo {author} {\bibfnamefont {A.}~\bibnamefont {Deller}}, \bibinfo
  {author} {\bibfnamefont {S.}~\bibnamefont {Eriksson}}, \bibinfo {author}
  {\bibfnamefont {J.}~\bibnamefont {Escallier}}, \bibinfo {author}
  {\bibfnamefont {J.}~\bibnamefont {Fajans}}, \bibinfo {author} {\bibfnamefont
  {T.}~\bibnamefont {Friesen}}, \bibinfo {author} {\bibfnamefont {M.~C.}\
  \bibnamefont {Fujiwara}}, \bibinfo {author} {\bibfnamefont {D.~R.}\
  \bibnamefont {Gill}}, \bibinfo {author} {\bibfnamefont {A.}~\bibnamefont
  {Gutierrez}}, \bibinfo {author} {\bibfnamefont {J.~S.}\ \bibnamefont
  {Hangst}}, \bibinfo {author} {\bibfnamefont {W.~N.}\ \bibnamefont {Hardy}},
  \bibinfo {author} {\bibfnamefont {R.~S.}\ \bibnamefont {Hayano}}, \bibinfo
  {author} {\bibfnamefont {M.~E.}\ \bibnamefont {Hayden}}, \bibinfo {author}
  {\bibfnamefont {A.~J.}\ \bibnamefont {Humphries}}, \bibinfo {author}
  {\bibfnamefont {J.~L.}\ \bibnamefont {Hurt}}, \bibinfo {author}
  {\bibfnamefont {R.}~\bibnamefont {Hydomako}}, \bibinfo {author}
  {\bibfnamefont {C.~A.}\ \bibnamefont {Isaac}}, \bibinfo {author}
  {\bibfnamefont {M.~J.}\ \bibnamefont {Jenkins}}, \bibinfo {author}
  {\bibfnamefont {S.}~\bibnamefont {Jonsell}}, \bibinfo {author} {\bibfnamefont
  {L.~V.}\ \bibnamefont {J??rgensen}}, \bibinfo {author} {\bibfnamefont
  {S.~J.}\ \bibnamefont {Kerrigan}}, \bibinfo {author} {\bibfnamefont
  {L.}~\bibnamefont {Kurchaninov}}, \bibinfo {author} {\bibfnamefont
  {N.}~\bibnamefont {Madsen}}, \bibinfo {author} {\bibfnamefont
  {A.}~\bibnamefont {Marone}}, \bibinfo {author} {\bibfnamefont {J.~T.~K.}\
  \bibnamefont {McKenna}}, \bibinfo {author} {\bibfnamefont {S.}~\bibnamefont
  {Menary}}, \bibinfo {author} {\bibfnamefont {P.}~\bibnamefont {Nolan}},
  \bibinfo {author} {\bibfnamefont {K.}~\bibnamefont {Olchanski}}, \bibinfo
  {author} {\bibfnamefont {A.}~\bibnamefont {Olin}}, \bibinfo {author}
  {\bibfnamefont {B.}~\bibnamefont {Parker}}, \bibinfo {author} {\bibfnamefont
  {A.}~\bibnamefont {Povilus}}, \bibinfo {author} {\bibfnamefont
  {P.}~\bibnamefont {Pusa}}, \bibinfo {author} {\bibfnamefont {F.}~\bibnamefont
  {Robicheaux}}, \bibinfo {author} {\bibfnamefont {E.}~\bibnamefont {Sarid}},
  \bibinfo {author} {\bibfnamefont {D.}~\bibnamefont {Seddon}}, \bibinfo
  {author} {\bibfnamefont {S.}~\bibnamefont {{Seif El Nasr}}}, \bibinfo
  {author} {\bibfnamefont {D.~M.}\ \bibnamefont {Silveira}}, \bibinfo {author}
  {\bibfnamefont {C.}~\bibnamefont {So}}, \bibinfo {author} {\bibfnamefont
  {J.~W.}\ \bibnamefont {Storey}}, \bibinfo {author} {\bibfnamefont {R.~I.}\
  \bibnamefont {Thompson}}, \bibinfo {author} {\bibfnamefont {J.}~\bibnamefont
  {Thornhill}}, \bibinfo {author} {\bibfnamefont {D.}~\bibnamefont {Wells}},
  \bibinfo {author} {\bibfnamefont {D.~P.}\ \bibnamefont {van~der Werf}},
  \bibinfo {author} {\bibfnamefont {J.~S.}\ \bibnamefont {Wurtele}}, \ and\
  \bibinfo {author} {\bibfnamefont {Y.}~\bibnamefont {Yamazaki}},\ }\href
  {\doibase 10.1016/j.nima.2013.09.043} {\bibfield  {journal} {\bibinfo
  {journal} {Nuclear Instruments and Methods in Physics Research, Section A:
  Accelerators, Spectrometers, Detectors and Associated Equipment}\ }\textbf
  {\bibinfo {volume} {735}},\ \bibinfo {pages} {319} (\bibinfo {year}
  {2014}{\natexlab{a}})}\BibitemShut {NoStop}%
\bibitem [{\citenamefont {Amole}\ \emph
  {et~al.}(2014{\natexlab{b}})\citenamefont {Amole}, \citenamefont {Ashkezari},
  \citenamefont {Baquero-Ruiz}, \citenamefont {Bertsche}, \citenamefont
  {Butler}, \citenamefont {Capra}, \citenamefont {Cesar}, \citenamefont
  {Charlton}, \citenamefont {Eriksson}, \citenamefont {Fajans}, \citenamefont
  {Friesen}, \citenamefont {Fujiwara}, \citenamefont {Gill}, \citenamefont
  {Gutierrez}, \citenamefont {Hangst}, \citenamefont {Hardy}, \citenamefont
  {Hayden}, \citenamefont {Isaac}, \citenamefont {Jonsell}, \citenamefont
  {Kurchaninov}, \citenamefont {Little}, \citenamefont {Madsen}, \citenamefont
  {McKenna}, \citenamefont {Menary}, \citenamefont {Napoli}, \citenamefont
  {Nolan}, \citenamefont {Olchanski}, \citenamefont {Olin}, \citenamefont
  {Povilus}, \citenamefont {Pusa}, \citenamefont {Rasmussen}, \citenamefont
  {Robicheaux}, \citenamefont {Sarid}, \citenamefont {Silveira}, \citenamefont
  {So}, \citenamefont {Tharp}, \citenamefont {Thompson}, \citenamefont {van~der
  Werf}, \citenamefont {Vendeiro}, \citenamefont {Wurtele}, \citenamefont
  {Zhmoginov},\ and\ \citenamefont {Charman}}]{Amole2014a}%
  \BibitemOpen
  \bibfield  {author} {\bibinfo {author} {\bibfnamefont {C.}~\bibnamefont
  {Amole}}, \bibinfo {author} {\bibfnamefont {M.~D.}\ \bibnamefont
  {Ashkezari}}, \bibinfo {author} {\bibfnamefont {M.}~\bibnamefont
  {Baquero-Ruiz}}, \bibinfo {author} {\bibfnamefont {W.}~\bibnamefont
  {Bertsche}}, \bibinfo {author} {\bibfnamefont {E.}~\bibnamefont {Butler}},
  \bibinfo {author} {\bibfnamefont {A.}~\bibnamefont {Capra}}, \bibinfo
  {author} {\bibfnamefont {C.~L.}\ \bibnamefont {Cesar}}, \bibinfo {author}
  {\bibfnamefont {M.}~\bibnamefont {Charlton}}, \bibinfo {author}
  {\bibfnamefont {S.}~\bibnamefont {Eriksson}}, \bibinfo {author}
  {\bibfnamefont {J.}~\bibnamefont {Fajans}}, \bibinfo {author} {\bibfnamefont
  {T.}~\bibnamefont {Friesen}}, \bibinfo {author} {\bibfnamefont {M.~C.}\
  \bibnamefont {Fujiwara}}, \bibinfo {author} {\bibfnamefont {D.~R.}\
  \bibnamefont {Gill}}, \bibinfo {author} {\bibfnamefont {A.}~\bibnamefont
  {Gutierrez}}, \bibinfo {author} {\bibfnamefont {J.~S.}\ \bibnamefont
  {Hangst}}, \bibinfo {author} {\bibfnamefont {W.~N.}\ \bibnamefont {Hardy}},
  \bibinfo {author} {\bibfnamefont {M.~E.}\ \bibnamefont {Hayden}}, \bibinfo
  {author} {\bibfnamefont {C.~A.}\ \bibnamefont {Isaac}}, \bibinfo {author}
  {\bibfnamefont {S.}~\bibnamefont {Jonsell}}, \bibinfo {author} {\bibfnamefont
  {L.}~\bibnamefont {Kurchaninov}}, \bibinfo {author} {\bibfnamefont
  {A.}~\bibnamefont {Little}}, \bibinfo {author} {\bibfnamefont
  {N.}~\bibnamefont {Madsen}}, \bibinfo {author} {\bibfnamefont {J.~T.~K.}\
  \bibnamefont {McKenna}}, \bibinfo {author} {\bibfnamefont {S.}~\bibnamefont
  {Menary}}, \bibinfo {author} {\bibfnamefont {S.~C.}\ \bibnamefont {Napoli}},
  \bibinfo {author} {\bibfnamefont {P.}~\bibnamefont {Nolan}}, \bibinfo
  {author} {\bibfnamefont {K.}~\bibnamefont {Olchanski}}, \bibinfo {author}
  {\bibfnamefont {A.}~\bibnamefont {Olin}}, \bibinfo {author} {\bibfnamefont
  {A.}~\bibnamefont {Povilus}}, \bibinfo {author} {\bibfnamefont
  {P.}~\bibnamefont {Pusa}}, \bibinfo {author} {\bibfnamefont {C.~{\O}.}\
  \bibnamefont {Rasmussen}}, \bibinfo {author} {\bibfnamefont {F.}~\bibnamefont
  {Robicheaux}}, \bibinfo {author} {\bibfnamefont {E.}~\bibnamefont {Sarid}},
  \bibinfo {author} {\bibfnamefont {D.~M.}\ \bibnamefont {Silveira}}, \bibinfo
  {author} {\bibfnamefont {C.}~\bibnamefont {So}}, \bibinfo {author}
  {\bibfnamefont {T.~D.}\ \bibnamefont {Tharp}}, \bibinfo {author}
  {\bibfnamefont {R.~I.}\ \bibnamefont {Thompson}}, \bibinfo {author}
  {\bibfnamefont {D.~P.}\ \bibnamefont {van~der Werf}}, \bibinfo {author}
  {\bibfnamefont {Z.}~\bibnamefont {Vendeiro}}, \bibinfo {author}
  {\bibfnamefont {J.~S.}\ \bibnamefont {Wurtele}}, \bibinfo {author}
  {\bibfnamefont {A.~I.}\ \bibnamefont {Zhmoginov}}, \ and\ \bibinfo {author}
  {\bibfnamefont {A.~E.}\ \bibnamefont {Charman}},\ }\href {\doibase
  10.1038/ncomms4955} {\bibfield  {journal} {\bibinfo  {journal} {Nature
  Communications}\ }\textbf {\bibinfo {volume} {5}},\ \bibinfo {pages} {3955}
  (\bibinfo {year} {2014}{\natexlab{b}})}\BibitemShut {NoStop}%
\bibitem [{\citenamefont {Mohri}\ and\ \citenamefont
  {Yamazaki}(2003)}]{Mohri2003}%
  \BibitemOpen
  \bibfield  {author} {\bibinfo {author} {\bibfnamefont {A.}~\bibnamefont
  {Mohri}}\ and\ \bibinfo {author} {\bibfnamefont {Y.}~\bibnamefont
  {Yamazaki}},\ }\href {\doibase 10.1209/epl/i2003-00509-0} {\bibfield
  {journal} {\bibinfo  {journal} {Europhysics Letters (EPL)}\ }\textbf
  {\bibinfo {volume} {63}},\ \bibinfo {pages} {207} (\bibinfo {year}
  {2003})}\BibitemShut {NoStop}%
\bibitem [{\citenamefont {Enomoto}\ \emph {et~al.}(2010)\citenamefont
  {Enomoto}, \citenamefont {Kuroda}, \citenamefont {Michishio}, \citenamefont
  {Kim}, \citenamefont {Higaki}, \citenamefont {Nagata}, \citenamefont {Kanai},
  \citenamefont {Torii}, \citenamefont {Corradini}, \citenamefont {Leali},
  \citenamefont {Lodi-Rizzini}, \citenamefont {Mascagna}, \citenamefont
  {Venturelli}, \citenamefont {Zurlo}, \citenamefont {Fujii}, \citenamefont
  {Ohtsuka}, \citenamefont {Tanaka}, \citenamefont {Imao}, \citenamefont
  {Nagashima}, \citenamefont {Matsuda}, \citenamefont {Juh{\'{a}}sz},
  \citenamefont {Mohri},\ and\ \citenamefont {Yamazaki}}]{Enomoto2010}%
  \BibitemOpen
  \bibfield  {author} {\bibinfo {author} {\bibfnamefont {Y.}~\bibnamefont
  {Enomoto}}, \bibinfo {author} {\bibfnamefont {N.}~\bibnamefont {Kuroda}},
  \bibinfo {author} {\bibfnamefont {K.}~\bibnamefont {Michishio}}, \bibinfo
  {author} {\bibfnamefont {C.~H.}\ \bibnamefont {Kim}}, \bibinfo {author}
  {\bibfnamefont {H.}~\bibnamefont {Higaki}}, \bibinfo {author} {\bibfnamefont
  {Y.}~\bibnamefont {Nagata}}, \bibinfo {author} {\bibfnamefont
  {Y.}~\bibnamefont {Kanai}}, \bibinfo {author} {\bibfnamefont {H.~A.}\
  \bibnamefont {Torii}}, \bibinfo {author} {\bibfnamefont {M.}~\bibnamefont
  {Corradini}}, \bibinfo {author} {\bibfnamefont {M.}~\bibnamefont {Leali}},
  \bibinfo {author} {\bibfnamefont {E.}~\bibnamefont {Lodi-Rizzini}}, \bibinfo
  {author} {\bibfnamefont {V.}~\bibnamefont {Mascagna}}, \bibinfo {author}
  {\bibfnamefont {L.}~\bibnamefont {Venturelli}}, \bibinfo {author}
  {\bibfnamefont {N.}~\bibnamefont {Zurlo}}, \bibinfo {author} {\bibfnamefont
  {K.}~\bibnamefont {Fujii}}, \bibinfo {author} {\bibfnamefont
  {M.}~\bibnamefont {Ohtsuka}}, \bibinfo {author} {\bibfnamefont
  {K.}~\bibnamefont {Tanaka}}, \bibinfo {author} {\bibfnamefont
  {H.}~\bibnamefont {Imao}}, \bibinfo {author} {\bibfnamefont {Y.}~\bibnamefont
  {Nagashima}}, \bibinfo {author} {\bibfnamefont {Y.}~\bibnamefont {Matsuda}},
  \bibinfo {author} {\bibfnamefont {B.}~\bibnamefont {Juh{\'{a}}sz}}, \bibinfo
  {author} {\bibfnamefont {A.}~\bibnamefont {Mohri}}, \ and\ \bibinfo {author}
  {\bibfnamefont {Y.}~\bibnamefont {Yamazaki}},\ }\href {\doibase
  10.1103/PhysRevLett.105.243401} {\bibfield  {journal} {\bibinfo  {journal}
  {Physical Review Letters}\ }\textbf {\bibinfo {volume} {105}},\ \bibinfo
  {pages} {1} (\bibinfo {year} {2010})}\BibitemShut {NoStop}%
\bibitem [{\citenamefont {Kuroda}\ \emph {et~al.}(2014)\citenamefont {Kuroda},
  \citenamefont {Ulmer}, \citenamefont {Murtagh}, \citenamefont {{Van Gorp}},
  \citenamefont {Nagata}, \citenamefont {Diermaier}, \citenamefont {Federmann},
  \citenamefont {Leali}, \citenamefont {Malbrunot}, \citenamefont {Mascagna},
  \citenamefont {Massiczek}, \citenamefont {Michishio}, \citenamefont
  {Mizutani}, \citenamefont {Mohri}, \citenamefont {Nagahama}, \citenamefont
  {Ohtsuka}, \citenamefont {Radics}, \citenamefont {Sakurai}, \citenamefont
  {Sauerzopf}, \citenamefont {Suzuki}, \citenamefont {Tajima}, \citenamefont
  {Torii}, \citenamefont {Venturelli}, \citenamefont {W{\"{u}}nschek},
  \citenamefont {Zmeskal}, \citenamefont {Zurlo}, \citenamefont {Higaki},
  \citenamefont {Kanai}, \citenamefont {Lodi-Rizzini}, \citenamefont
  {Nagashima}, \citenamefont {Matsuda}, \citenamefont {Widmann},\ and\
  \citenamefont {Yamazaki}}]{Kuroda2014}%
  \BibitemOpen
  \bibfield  {author} {\bibinfo {author} {\bibfnamefont {N.}~\bibnamefont
  {Kuroda}}, \bibinfo {author} {\bibfnamefont {S.}~\bibnamefont {Ulmer}},
  \bibinfo {author} {\bibfnamefont {D.~J.}\ \bibnamefont {Murtagh}}, \bibinfo
  {author} {\bibfnamefont {S.}~\bibnamefont {{Van Gorp}}}, \bibinfo {author}
  {\bibfnamefont {Y.}~\bibnamefont {Nagata}}, \bibinfo {author} {\bibfnamefont
  {M.}~\bibnamefont {Diermaier}}, \bibinfo {author} {\bibfnamefont
  {S.}~\bibnamefont {Federmann}}, \bibinfo {author} {\bibfnamefont
  {M.}~\bibnamefont {Leali}}, \bibinfo {author} {\bibfnamefont
  {C.}~\bibnamefont {Malbrunot}}, \bibinfo {author} {\bibfnamefont
  {V.}~\bibnamefont {Mascagna}}, \bibinfo {author} {\bibfnamefont
  {O.}~\bibnamefont {Massiczek}}, \bibinfo {author} {\bibfnamefont
  {K.}~\bibnamefont {Michishio}}, \bibinfo {author} {\bibfnamefont
  {T.}~\bibnamefont {Mizutani}}, \bibinfo {author} {\bibfnamefont
  {a.}~\bibnamefont {Mohri}}, \bibinfo {author} {\bibfnamefont
  {H.}~\bibnamefont {Nagahama}}, \bibinfo {author} {\bibfnamefont
  {M.}~\bibnamefont {Ohtsuka}}, \bibinfo {author} {\bibfnamefont
  {B.}~\bibnamefont {Radics}}, \bibinfo {author} {\bibfnamefont
  {S.}~\bibnamefont {Sakurai}}, \bibinfo {author} {\bibfnamefont
  {C.}~\bibnamefont {Sauerzopf}}, \bibinfo {author} {\bibfnamefont
  {K.}~\bibnamefont {Suzuki}}, \bibinfo {author} {\bibfnamefont
  {M.}~\bibnamefont {Tajima}}, \bibinfo {author} {\bibfnamefont {H.~A.}\
  \bibnamefont {Torii}}, \bibinfo {author} {\bibfnamefont {L.}~\bibnamefont
  {Venturelli}}, \bibinfo {author} {\bibfnamefont {B.}~\bibnamefont
  {W{\"{u}}nschek}}, \bibinfo {author} {\bibfnamefont {J.}~\bibnamefont
  {Zmeskal}}, \bibinfo {author} {\bibfnamefont {N.}~\bibnamefont {Zurlo}},
  \bibinfo {author} {\bibfnamefont {H.}~\bibnamefont {Higaki}}, \bibinfo
  {author} {\bibfnamefont {Y.}~\bibnamefont {Kanai}}, \bibinfo {author}
  {\bibfnamefont {E.}~\bibnamefont {Lodi-Rizzini}}, \bibinfo {author}
  {\bibfnamefont {Y.}~\bibnamefont {Nagashima}}, \bibinfo {author}
  {\bibfnamefont {Y.}~\bibnamefont {Matsuda}}, \bibinfo {author} {\bibfnamefont
  {E.}~\bibnamefont {Widmann}}, \ and\ \bibinfo {author} {\bibfnamefont
  {Y.}~\bibnamefont {Yamazaki}},\ }\href {\doibase 10.1038/ncomms4089}
  {\bibfield  {journal} {\bibinfo  {journal} {Nature communications}\ }\textbf
  {\bibinfo {volume} {5}},\ \bibinfo {pages} {3089} (\bibinfo {year}
  {2014})}\BibitemShut {NoStop}%
\bibitem [{\citenamefont {Walz}\ \emph {et~al.}(1995)\citenamefont {Walz},
  \citenamefont {Ross}, \citenamefont {Zimmermann}, \citenamefont {Ricci},
  \citenamefont {Prevedelli},\ and\ \citenamefont {H{\"{a}}nsch}}]{Walz1995}%
  \BibitemOpen
  \bibfield  {author} {\bibinfo {author} {\bibfnamefont {J.}~\bibnamefont
  {Walz}}, \bibinfo {author} {\bibfnamefont {S.}~\bibnamefont {Ross}}, \bibinfo
  {author} {\bibfnamefont {C.}~\bibnamefont {Zimmermann}}, \bibinfo {author}
  {\bibfnamefont {L.}~\bibnamefont {Ricci}}, \bibinfo {author} {\bibfnamefont
  {M.}~\bibnamefont {Prevedelli}}, \ and\ \bibinfo {author} {\bibfnamefont
  {T.~W.}\ \bibnamefont {H{\"{a}}nsch}},\ }\href {\doibase
  10.1103/PhysRevLett.75.3257} {\bibfield  {journal} {\bibinfo  {journal}
  {Physical Review Letters}\ }\textbf {\bibinfo {volume} {75}},\ \bibinfo
  {pages} {3257} (\bibinfo {year} {1995})}\BibitemShut {NoStop}%
\bibitem [{\citenamefont {Hoffrogge}\ \emph {et~al.}(2011)\citenamefont
  {Hoffrogge}, \citenamefont {Fr{\"{o}}hlich}, \citenamefont {Kasevich},\ and\
  \citenamefont {Hommelhoff}}]{Hoffrogge2011}%
  \BibitemOpen
  \bibfield  {author} {\bibinfo {author} {\bibfnamefont {J.}~\bibnamefont
  {Hoffrogge}}, \bibinfo {author} {\bibfnamefont {R.}~\bibnamefont
  {Fr{\"{o}}hlich}}, \bibinfo {author} {\bibfnamefont {M.~A.}\ \bibnamefont
  {Kasevich}}, \ and\ \bibinfo {author} {\bibfnamefont {P.}~\bibnamefont
  {Hommelhoff}},\ }\href {\doibase 10.1103/PhysRevLett.106.193001} {\bibfield
  {journal} {\bibinfo  {journal} {Physical Review Letters}\ }\textbf {\bibinfo
  {volume} {106}},\ \bibinfo {pages} {1} (\bibinfo {year} {2011})},\ \Eprint
  {http://arxiv.org/abs/1012.2376} {arXiv:1012.2376} \BibitemShut {NoStop}%
\bibitem [{\citenamefont {Werth}\ \emph {et~al.}(2009)\citenamefont {Werth},
  \citenamefont {Gheorghe},\ and\ \citenamefont {Major}}]{Seriesa}%
  \BibitemOpen
  \bibfield  {author} {\bibinfo {author} {\bibfnamefont {G.}~\bibnamefont
  {Werth}}, \bibinfo {author} {\bibfnamefont {V.~N.}\ \bibnamefont {Gheorghe}},
  \ and\ \bibinfo {author} {\bibfnamefont {F.~G.}\ \bibnamefont {Major}},\
  }\href {\doibase 10.1007/978-3-540-92261-2} {\emph {\bibinfo {title}
  {{Charged Particle Traps II}}}},\ \bibinfo {series} {Springer Series on
  Atomic, Optical, and Plasma Physics}, Vol.~\bibinfo {volume} {54}\ (\bibinfo
  {publisher} {Springer Berlin Heidelberg},\ \bibinfo {address} {Berlin,
  Heidelberg},\ \bibinfo {year} {2009})\ p.\ \bibinfo {pages} {354}\BibitemShut
  {NoStop}%
\bibitem [{\citenamefont {Goldman}\ and\ \citenamefont
  {Dalibard}(2014)}]{Goldman2014}%
  \BibitemOpen
  \bibfield  {author} {\bibinfo {author} {\bibfnamefont {N.}~\bibnamefont
  {Goldman}}\ and\ \bibinfo {author} {\bibfnamefont {J.}~\bibnamefont
  {Dalibard}},\ }\href {\doibase 10.1103/PhysRevX.4.031027} {\bibfield
  {journal} {\bibinfo  {journal} {Phys. Rev. X}\ }\textbf {\bibinfo {volume}
  {4}},\ \bibinfo {pages} {31027} (\bibinfo {year} {2014})}\BibitemShut
  {NoStop}%
\bibitem [{\citenamefont {{ASACUSA Collaboration}}(2007)}]{ASACUSA2007}%
  \BibitemOpen
  \bibfield  {author} {\bibinfo {author} {\bibnamefont {{ASACUSA
  Collaboration}}},\ }\href@noop {} {\bibfield  {journal} {\bibinfo  {journal}
  {CERN-SPSC-2007-003}\ }\textbf {\bibinfo {volume} {SPSC-SR-01}} (\bibinfo
  {year} {2007})}\BibitemShut {NoStop}%
\bibitem [{\citenamefont {{ASACUSA Collaboration}}(2008)}]{ASACUSA2008}%
  \BibitemOpen
  \bibfield  {author} {\bibinfo {author} {\bibnamefont {{ASACUSA
  Collaboration}}},\ }\href
  {http://asacusa.web.cern.ch/ASACUSA/home/spsc/spsc-jan2002.pdf} {\bibfield
  {journal} {\bibinfo  {journal} {CERN-SPSC-2008-002}\ }\textbf {\bibinfo
  {volume} {SPSC-SR-02}} (\bibinfo {year} {2008})}\BibitemShut {NoStop}%
\bibitem [{\citenamefont {{ASACUSA Collaboration}}(2009)}]{ASACUSA2009}%
  \BibitemOpen
  \bibfield  {author} {\bibinfo {author} {\bibnamefont {{ASACUSA
  Collaboration}}},\ }\href@noop {} {\bibfield  {journal} {\bibinfo  {journal}
  {CERN-SPSC-2009-005}\ }\textbf {\bibinfo {volume} {SPSC-SR-04}} (\bibinfo
  {year} {2009})}\BibitemShut {NoStop}%
\bibitem [{\citenamefont {{ASACUSA Collaboration}}(2010)}]{ASACUSA2010}%
  \BibitemOpen
  \bibfield  {author} {\bibinfo {author} {\bibnamefont {{ASACUSA
  Collaboration}}},\ }\href@noop {} {\bibfield  {journal} {\bibinfo  {journal}
  {CERN-SPSC-2010-005}\ }\textbf {\bibinfo {volume} {SPSC-SR-05}} (\bibinfo
  {year} {2010})}\BibitemShut {NoStop}%
\bibitem [{\citenamefont {{ASACUSA Collaboration}}(2011)}]{ASACUSA2011}%
  \BibitemOpen
  \bibfield  {author} {\bibinfo {author} {\bibnamefont {{ASACUSA
  Collaboration}}},\ }\href@noop {} {\bibfield  {journal} {\bibinfo  {journal}
  {CERN-SPSC-2011-003}\ }\textbf {\bibinfo {volume} {SPSC-SR-07}} (\bibinfo
  {year} {2011})}\BibitemShut {NoStop}%
\bibitem [{\citenamefont {{ASACUSA Collaboration}}(2014)}]{ASACUSA2014}%
  \BibitemOpen
  \bibfield  {author} {\bibinfo {author} {\bibnamefont {{ASACUSA
  Collaboration}}},\ }\href
  {http://asacusa.web.cern.ch/ASACUSA/home/spsc/spsc-jan2002.pdf} {\bibfield
  {journal} {\bibinfo  {journal} {CERN-SPSC-2014-004}\ }\textbf {\bibinfo
  {volume} {CERN-SPSC-}} (\bibinfo {year} {2014})}\BibitemShut {NoStop}%
\bibitem [{\citenamefont {Reichel}\ and\ \citenamefont
  {Vuleti{\'{c}}}(2011)}]{Reichel2011}%
  \BibitemOpen
  \bibfield  {author} {\bibinfo {author} {\bibfnamefont {J.}~\bibnamefont
  {Reichel}}\ and\ \bibinfo {author} {\bibfnamefont {V.}~\bibnamefont
  {Vuleti{\'{c}}}},\ }\href {\doibase 10.1002/9783527633357} {\emph {\bibinfo
  {title} {Atom Chips}}},\ edited by\ \bibinfo {editor} {\bibfnamefont
  {J.}~\bibnamefont {Reichel}}\ and\ \bibinfo {editor} {\bibfnamefont
  {V.}~\bibnamefont {Vuleti{\'{c}}}}\ (\bibinfo  {publisher} {Wiley-VCH Verlag
  GmbH {\{}{\&}{\}} Co. KGaA},\ \bibinfo {address} {Weinheim, Germany},\
  \bibinfo {year} {2011})\BibitemShut {NoStop}%
\bibitem [{\citenamefont {Keil}\ \emph {et~al.}(2016)\citenamefont {Keil},
  \citenamefont {Amit}, \citenamefont {Zhou}, \citenamefont {Groswasser},
  \citenamefont {Japha},\ and\ \citenamefont {Folman}}]{Keil2016}%
  \BibitemOpen
  \bibfield  {author} {\bibinfo {author} {\bibfnamefont {M.}~\bibnamefont
  {Keil}}, \bibinfo {author} {\bibfnamefont {O.}~\bibnamefont {Amit}}, \bibinfo
  {author} {\bibfnamefont {S.}~\bibnamefont {Zhou}}, \bibinfo {author}
  {\bibfnamefont {D.}~\bibnamefont {Groswasser}}, \bibinfo {author}
  {\bibfnamefont {Y.}~\bibnamefont {Japha}}, \ and\ \bibinfo {author}
  {\bibfnamefont {R.}~\bibnamefont {Folman}},\ }\href@noop {} {\bibfield
  {journal} {\bibinfo  {journal} {J. Mod. Optics}\ } (\bibinfo {year}
  {2016})}\BibitemShut {NoStop}%
\bibitem [{\citenamefont {Trypogeorgos}\ and\ \citenamefont
  {Foot}(2013)}]{Trypogeorgos2013}%
  \BibitemOpen
  \bibfield  {author} {\bibinfo {author} {\bibfnamefont {D.}~\bibnamefont
  {Trypogeorgos}}\ and\ \bibinfo {author} {\bibfnamefont {C.}~\bibnamefont
  {Foot}},\ }\href {http://arxiv.org/abs/1310.6294} {\bibfield  {journal}
  {\bibinfo  {journal} {arXiv preprint arXiv:1310.6294}\ } (\bibinfo {year}
  {2013})},\ \Eprint {http://arxiv.org/abs/1310.6294} {arXiv:1310.6294}
  \BibitemShut {NoStop}%
\bibitem [{\citenamefont {Kaufmann}\ \emph {et~al.}(2012)\citenamefont
  {Kaufmann}, \citenamefont {Ulm}, \citenamefont {Jacob}, \citenamefont
  {Poschinger}, \citenamefont {Landa}, \citenamefont {Retzker}, \citenamefont
  {Plenio},\ and\ \citenamefont {Schmidt-Kaler}}]{Kaufmann2012}%
  \BibitemOpen
  \bibfield  {author} {\bibinfo {author} {\bibfnamefont {H.}~\bibnamefont
  {Kaufmann}}, \bibinfo {author} {\bibfnamefont {S.}~\bibnamefont {Ulm}},
  \bibinfo {author} {\bibfnamefont {G.}~\bibnamefont {Jacob}}, \bibinfo
  {author} {\bibfnamefont {U.}~\bibnamefont {Poschinger}}, \bibinfo {author}
  {\bibfnamefont {H.}~\bibnamefont {Landa}}, \bibinfo {author} {\bibfnamefont
  {A.}~\bibnamefont {Retzker}}, \bibinfo {author} {\bibfnamefont {M.~B.}\
  \bibnamefont {Plenio}}, \ and\ \bibinfo {author} {\bibfnamefont
  {F.}~\bibnamefont {Schmidt-Kaler}},\ }\href {\doibase
  10.1103/PhysRevLett.109.263003} {\bibfield  {journal} {\bibinfo  {journal}
  {Phys. Rev. Lett.}\ }\textbf {\bibinfo {volume} {109}},\ \bibinfo {pages}
  {263003} (\bibinfo {year} {2012})}\BibitemShut {NoStop}%
\bibitem [{\citenamefont {Landa}\ \emph
  {et~al.}(2012{\natexlab{a}})\citenamefont {Landa}, \citenamefont {Drewsen},
  \citenamefont {Reznik},\ and\ \citenamefont {Retzker}}]{Landa2012a}%
  \BibitemOpen
  \bibfield  {author} {\bibinfo {author} {\bibfnamefont {H.}~\bibnamefont
  {Landa}}, \bibinfo {author} {\bibfnamefont {M.}~\bibnamefont {Drewsen}},
  \bibinfo {author} {\bibfnamefont {B.}~\bibnamefont {Reznik}}, \ and\ \bibinfo
  {author} {\bibfnamefont {A.}~\bibnamefont {Retzker}},\ }\href
  {http://stacks.iop.org/1367-2630/14/i=9/a=093023} {\bibfield  {journal}
  {\bibinfo  {journal} {New Journal of Physics}\ }\textbf {\bibinfo {volume}
  {14}},\ \bibinfo {pages} {93023} (\bibinfo {year}
  {2012}{\natexlab{a}})}\BibitemShut {NoStop}%
\bibitem [{\citenamefont {Jordan}\ and\ \citenamefont
  {Smith}(2007)}]{Jordan2007}%
  \BibitemOpen
  \bibfield  {author} {\bibinfo {author} {\bibfnamefont {D.~W.}\ \bibnamefont
  {Jordan}}\ and\ \bibinfo {author} {\bibfnamefont {P.}~\bibnamefont {Smith}},\
  }\href@noop {} {\emph {\bibinfo {title} {{Nonlinear Ordinary Differential
  Equations: An introduction for Scientists and Engineers}}}},\ \bibinfo
  {edition} {fourth edi}\ ed.\ (\bibinfo  {publisher} {Oxford University
  Press},\ \bibinfo {year} {2007})\ p.\ \bibinfo {pages} {540}\BibitemShut
  {NoStop}%
\bibitem [{\citenamefont {Landa}\ \emph
  {et~al.}(2012{\natexlab{b}})\citenamefont {Landa}, \citenamefont {Drewsen},
  \citenamefont {Reznik},\ and\ \citenamefont {Retzker}}]{Landa2012b}%
  \BibitemOpen
  \bibfield  {author} {\bibinfo {author} {\bibfnamefont {H.}~\bibnamefont
  {Landa}}, \bibinfo {author} {\bibfnamefont {M.}~\bibnamefont {Drewsen}},
  \bibinfo {author} {\bibfnamefont {B.}~\bibnamefont {Reznik}}, \ and\ \bibinfo
  {author} {\bibfnamefont {A.}~\bibnamefont {Retzker}},\ }\href
  {http://stacks.iop.org/1751-8121/45/i=45/a=455305} {\bibfield  {journal}
  {\bibinfo  {journal} {Journal of Physics A: Mathematical and Theoretical}\
  }\textbf {\bibinfo {volume} {45}},\ \bibinfo {pages} {455305} (\bibinfo
  {year} {2012}{\natexlab{b}})}\BibitemShut {NoStop}%
\bibitem [{\citenamefont {Maero}\ \emph {et~al.}(2012)\citenamefont {Maero},
  \citenamefont {Herfurth}, \citenamefont {Kluge}, \citenamefont {Schwarz},\
  and\ \citenamefont {Zwicknagel}}]{Maero2012}%
  \BibitemOpen
  \bibfield  {author} {\bibinfo {author} {\bibfnamefont {G.}~\bibnamefont
  {Maero}}, \bibinfo {author} {\bibfnamefont {F.}~\bibnamefont {Herfurth}},
  \bibinfo {author} {\bibfnamefont {H.~J.}\ \bibnamefont {Kluge}}, \bibinfo
  {author} {\bibfnamefont {S.}~\bibnamefont {Schwarz}}, \ and\ \bibinfo
  {author} {\bibfnamefont {G.}~\bibnamefont {Zwicknagel}},\ }\href {\doibase
  10.1007/s00340-011-4808-5} {\bibfield  {journal} {\bibinfo  {journal}
  {Applied Physics B: Lasers and Optics}\ }\textbf {\bibinfo {volume} {107}},\
  \bibinfo {pages} {1087} (\bibinfo {year} {2012})}\BibitemShut {NoStop}%
\bibitem [{\citenamefont {Ulmer}\ \emph {et~al.}(2013)\citenamefont {Ulmer},
  \citenamefont {Blaum}, \citenamefont {Kracke}, \citenamefont {Mooser},
  \citenamefont {Quint}, \citenamefont {Rodegheri},\ and\ \citenamefont
  {Walz}}]{Ulmer2013}%
  \BibitemOpen
  \bibfield  {author} {\bibinfo {author} {\bibfnamefont {S.}~\bibnamefont
  {Ulmer}}, \bibinfo {author} {\bibfnamefont {K.}~\bibnamefont {Blaum}},
  \bibinfo {author} {\bibfnamefont {H.}~\bibnamefont {Kracke}}, \bibinfo
  {author} {\bibfnamefont {A.}~\bibnamefont {Mooser}}, \bibinfo {author}
  {\bibfnamefont {W.}~\bibnamefont {Quint}}, \bibinfo {author} {\bibfnamefont
  {C.~C.}\ \bibnamefont {Rodegheri}}, \ and\ \bibinfo {author} {\bibfnamefont
  {J.}~\bibnamefont {Walz}},\ }\href {\doibase 10.1016/j.nima.2012.12.071}
  {\bibfield  {journal} {\bibinfo  {journal} {Nuclear Instruments and Methods
  in Physics Research Section A: Accelerators, Spectrometers, Detectors and
  Associated Equipment}\ }\textbf {\bibinfo {volume} {705}},\ \bibinfo {pages}
  {55} (\bibinfo {year} {2013})}\BibitemShut {NoStop}%
\bibitem [{\citenamefont {Geyer}\ and\ \citenamefont
  {Blumel}(2012)}]{Geyer2012}%
  \BibitemOpen
  \bibfield  {author} {\bibinfo {author} {\bibfnamefont {G.}~\bibnamefont
  {Geyer}}\ and\ \bibinfo {author} {\bibfnamefont {R.}~\bibnamefont {Blumel}},\
  }\href {http://www.jurp.org/2012/MS128.pdf} {\bibfield  {journal} {\bibinfo
  {journal} {Journal of Undergraduate Research in Physics}\ }\textbf {\bibinfo
  {volume} {25}},\ \bibinfo {pages} {1} (\bibinfo {year} {2012})}\BibitemShut
  {NoStop}%
\bibitem [{\citenamefont {Gabrielse}\ \emph {et~al.}(1989)\citenamefont
  {Gabrielse}, \citenamefont {Rolston}, \citenamefont {Haarsma},\ and\
  \citenamefont {Kells}}]{Gabrielse1989}%
  \BibitemOpen
  \bibfield  {author} {\bibinfo {author} {\bibfnamefont {G.}~\bibnamefont
  {Gabrielse}}, \bibinfo {author} {\bibfnamefont {S.~L.}\ \bibnamefont
  {Rolston}}, \bibinfo {author} {\bibfnamefont {L.}~\bibnamefont {Haarsma}}, \
  and\ \bibinfo {author} {\bibfnamefont {W.}~\bibnamefont {Kells}},\ }\href
  {\doibase 10.1007/BF02398677} {\bibfield  {journal} {\bibinfo  {journal}
  {Hyperfine Interactions}\ }\textbf {\bibinfo {volume} {44}},\ \bibinfo
  {pages} {287} (\bibinfo {year} {1989})}\BibitemShut {NoStop}%
\bibitem [{\citenamefont {Yousif}\ \emph {et~al.}(1991)\citenamefont {Yousif},
  \citenamefont {{Van der Donk}}, \citenamefont {Kucherovsky}, \citenamefont
  {Reis}, \citenamefont {Brannen}, \citenamefont {Mitchell},\ and\
  \citenamefont {Morgan}}]{Yousif1991}%
  \BibitemOpen
  \bibfield  {author} {\bibinfo {author} {\bibfnamefont {F.}~\bibnamefont
  {Yousif}}, \bibinfo {author} {\bibfnamefont {P.}~\bibnamefont {{Van der
  Donk}}}, \bibinfo {author} {\bibfnamefont {Z.}~\bibnamefont {Kucherovsky}},
  \bibinfo {author} {\bibfnamefont {J.}~\bibnamefont {Reis}}, \bibinfo {author}
  {\bibfnamefont {E.}~\bibnamefont {Brannen}}, \bibinfo {author} {\bibfnamefont
  {J.}~\bibnamefont {Mitchell}}, \ and\ \bibinfo {author} {\bibfnamefont
  {T.}~\bibnamefont {Morgan}},\ }\href {\doibase 10.1103/PhysRevLett.67.26}
  {\bibfield  {journal} {\bibinfo  {journal} {Physical Review Letters}\
  }\textbf {\bibinfo {volume} {67}},\ \bibinfo {pages} {26} (\bibinfo {year}
  {1991})}\BibitemShut {NoStop}%
\bibitem [{\citenamefont {Wolf}(1993)}]{Wolf1993}%
  \BibitemOpen
  \bibfield  {author} {\bibinfo {author} {\bibfnamefont {A.}~\bibnamefont
  {Wolf}},\ }\href {\doibase 10.1007/BF02316718} {\bibfield  {journal}
  {\bibinfo  {journal} {Hyperfine Interactions}\ }\textbf {\bibinfo {volume}
  {76}},\ \bibinfo {pages} {189} (\bibinfo {year} {1993})}\BibitemShut
  {NoStop}%
\bibitem [{\citenamefont {Amoretti}\ \emph {et~al.}(2006)\citenamefont
  {Amoretti}, \citenamefont {Amsler}, \citenamefont {Bonomi}, \citenamefont
  {Bowe}, \citenamefont {Canali}, \citenamefont {Carraro}, \citenamefont
  {Cesar}, \citenamefont {Charlton}, \citenamefont {Ejsing}, \citenamefont
  {Fontana}, \citenamefont {Fujiwara}, \citenamefont {Funakoshi}, \citenamefont
  {Genova}, \citenamefont {Hangst}, \citenamefont {Hayano}, \citenamefont
  {J{\o}rgensen}, \citenamefont {Kellerbauer}, \citenamefont {Lagomarsino},
  \citenamefont {Lodi-Rizzini}, \citenamefont {MacR$\backslash$`$\backslash$i},
  \citenamefont {Madsen}, \citenamefont {Manuzio}, \citenamefont {Mitchard},
  \citenamefont {Montagna}, \citenamefont {Posada}, \citenamefont {Pruys},
  \citenamefont {Regenfus}, \citenamefont {Rotondi}, \citenamefont {Telle},
  \citenamefont {Testera}, \citenamefont {van~der Werf}, \citenamefont
  {Variola}, \citenamefont {Venturelli}, \citenamefont {Yamazaki},\ and\
  \citenamefont {Zurlo}}]{Amoretti2006}%
  \BibitemOpen
  \bibfield  {author} {\bibinfo {author} {\bibfnamefont {M.}~\bibnamefont
  {Amoretti}}, \bibinfo {author} {\bibfnamefont {C.}~\bibnamefont {Amsler}},
  \bibinfo {author} {\bibfnamefont {G.}~\bibnamefont {Bonomi}}, \bibinfo
  {author} {\bibfnamefont {P.~D.}\ \bibnamefont {Bowe}}, \bibinfo {author}
  {\bibfnamefont {C.}~\bibnamefont {Canali}}, \bibinfo {author} {\bibfnamefont
  {C.}~\bibnamefont {Carraro}}, \bibinfo {author} {\bibfnamefont {C.~L.}\
  \bibnamefont {Cesar}}, \bibinfo {author} {\bibfnamefont {M.}~\bibnamefont
  {Charlton}}, \bibinfo {author} {\bibfnamefont {a.~M.}\ \bibnamefont
  {Ejsing}}, \bibinfo {author} {\bibfnamefont {A.}~\bibnamefont {Fontana}},
  \bibinfo {author} {\bibfnamefont {M.~C.}\ \bibnamefont {Fujiwara}}, \bibinfo
  {author} {\bibfnamefont {R.}~\bibnamefont {Funakoshi}}, \bibinfo {author}
  {\bibfnamefont {P.}~\bibnamefont {Genova}}, \bibinfo {author} {\bibfnamefont
  {J.~S.}\ \bibnamefont {Hangst}}, \bibinfo {author} {\bibfnamefont {R.~S.}\
  \bibnamefont {Hayano}}, \bibinfo {author} {\bibfnamefont {L.~V.}\
  \bibnamefont {J{\o}rgensen}}, \bibinfo {author} {\bibfnamefont
  {A.}~\bibnamefont {Kellerbauer}}, \bibinfo {author} {\bibfnamefont
  {V.}~\bibnamefont {Lagomarsino}}, \bibinfo {author} {\bibfnamefont
  {E.}~\bibnamefont {Lodi-Rizzini}}, \bibinfo {author} {\bibfnamefont
  {M.}~\bibnamefont {MacR$\backslash$`$\backslash$i}}, \bibinfo {author}
  {\bibfnamefont {N.}~\bibnamefont {Madsen}}, \bibinfo {author} {\bibfnamefont
  {G.}~\bibnamefont {Manuzio}}, \bibinfo {author} {\bibfnamefont
  {D.}~\bibnamefont {Mitchard}}, \bibinfo {author} {\bibfnamefont
  {P.}~\bibnamefont {Montagna}}, \bibinfo {author} {\bibfnamefont {L.~G.~C.}\
  \bibnamefont {Posada}}, \bibinfo {author} {\bibfnamefont {H.}~\bibnamefont
  {Pruys}}, \bibinfo {author} {\bibfnamefont {C.}~\bibnamefont {Regenfus}},
  \bibinfo {author} {\bibfnamefont {A.}~\bibnamefont {Rotondi}}, \bibinfo
  {author} {\bibfnamefont {H.~H.}\ \bibnamefont {Telle}}, \bibinfo {author}
  {\bibfnamefont {G.}~\bibnamefont {Testera}}, \bibinfo {author} {\bibfnamefont
  {D.~P.}\ \bibnamefont {van~der Werf}}, \bibinfo {author} {\bibfnamefont
  {a.}~\bibnamefont {Variola}}, \bibinfo {author} {\bibfnamefont
  {L.}~\bibnamefont {Venturelli}}, \bibinfo {author} {\bibfnamefont
  {Y.}~\bibnamefont {Yamazaki}}, \ and\ \bibinfo {author} {\bibfnamefont
  {N.}~\bibnamefont {Zurlo}},\ }\href {\doibase 10.1103/PhysRevLett.97.213401}
  {\bibfield  {journal} {\bibinfo  {journal} {Physical Review Letters}\
  }\textbf {\bibinfo {volume} {97}},\ \bibinfo {pages} {1} (\bibinfo {year}
  {2006})}\BibitemShut {NoStop}%
\bibitem [{\citenamefont {Kolbe}\ \emph {et~al.}(2011)\citenamefont {Kolbe},
  \citenamefont {Beczkowiak}, \citenamefont {Diehl}, \citenamefont {Koglbauer},
  \citenamefont {Sattler}, \citenamefont {Stappel}, \citenamefont {Steinborn},\
  and\ \citenamefont {Walz}}]{Kolbe2011}%
  \BibitemOpen
  \bibfield  {author} {\bibinfo {author} {\bibfnamefont {D.}~\bibnamefont
  {Kolbe}}, \bibinfo {author} {\bibfnamefont {A.}~\bibnamefont {Beczkowiak}},
  \bibinfo {author} {\bibfnamefont {T.}~\bibnamefont {Diehl}}, \bibinfo
  {author} {\bibfnamefont {A.}~\bibnamefont {Koglbauer}}, \bibinfo {author}
  {\bibfnamefont {M.}~\bibnamefont {Sattler}}, \bibinfo {author} {\bibfnamefont
  {M.}~\bibnamefont {Stappel}}, \bibinfo {author} {\bibfnamefont
  {R.}~\bibnamefont {Steinborn}}, \ and\ \bibinfo {author} {\bibfnamefont
  {J.}~\bibnamefont {Walz}},\ }\href {\doibase 10.1007/s10751-011-0381-x}
  {\bibfield  {journal} {\bibinfo  {journal} {Hyperfine Interactions}\ }\textbf
  {\bibinfo {volume} {212}},\ \bibinfo {pages} {213} (\bibinfo {year}
  {2011})}\BibitemShut {NoStop}%
\bibitem [{\citenamefont {Michan}\ \emph {et~al.}(2014)\citenamefont {Michan},
  \citenamefont {Fujiwara},\ and\ \citenamefont {Momose}}]{Michan2014}%
  \BibitemOpen
  \bibfield  {author} {\bibinfo {author} {\bibfnamefont {J.~M.}\ \bibnamefont
  {Michan}}, \bibinfo {author} {\bibfnamefont {M.~C.}\ \bibnamefont
  {Fujiwara}}, \ and\ \bibinfo {author} {\bibfnamefont {T.}~\bibnamefont
  {Momose}},\ }\href {\doibase 10.1007/s10751-014-1017-8} {\bibfield  {journal}
  {\bibinfo  {journal} {Hyperfine Interactions}\ }\textbf {\bibinfo {volume}
  {228}},\ \bibinfo {pages} {77} (\bibinfo {year} {2014})}\BibitemShut
  {NoStop}%
\bibitem [{\citenamefont {Alheit}\ \emph {et~al.}(1996)\citenamefont {Alheit},
  \citenamefont {Gudjons}, \citenamefont {Kleineidam},\ and\ \citenamefont
  {Werth}}]{Alheit1996a}%
  \BibitemOpen
  \bibfield  {author} {\bibinfo {author} {\bibfnamefont {R.}~\bibnamefont
  {Alheit}}, \bibinfo {author} {\bibfnamefont {T.}~\bibnamefont {Gudjons}},
  \bibinfo {author} {\bibfnamefont {S.}~\bibnamefont {Kleineidam}}, \ and\
  \bibinfo {author} {\bibfnamefont {G.}~\bibnamefont {Werth}},\ }\href
  {\doibase 10.1002/(SICI)1097-0231(19960331)10:5<583::AID-RCM497>3.0.CO;2-2}
  {\bibfield  {journal} {\bibinfo  {journal} {Rapid Communications in Mass
  Spectrometry}\ }\textbf {\bibinfo {volume} {10}},\ \bibinfo {pages} {583}
  (\bibinfo {year} {1996})}\BibitemShut {NoStop}%
\bibitem [{\citenamefont {{ASACUSA Collaboration}}(2005)}]{ASACUSA2005}%
  \BibitemOpen
  \bibfield  {author} {\bibinfo {author} {\bibnamefont {{ASACUSA
  Collaboration}}},\ }\href@noop {} {\bibfield  {journal} {\bibinfo  {journal}
  {CERN-SPSC 2005-002}\ }\textbf {\bibinfo {volume} {SPSC P-307}} (\bibinfo
  {year} {2005})}\BibitemShut {NoStop}%
\end{thebibliography}%

\end{document}